\documentclass[a4paper,fleqn,usenatbib,useAMS]{mnras}
\usepackage{graphicx}   
\usepackage{color}
\usepackage{lscape}
\usepackage{txfonts}
\usepackage{amsfonts}

\title[Anti-correlation between $\Gamma$ and $L_{\rm 0.5-10keV}$]
{The advection-dominated accretion flow for the anti-correlation between the X-ray photon index 
and the X-ray luminosity in neutron star low-mass X-ray binaries}
\author[Erlin Qiao and B.F. Liu]{Erlin Qiao $^{1,2}$\thanks{E-mail:
qiaoel@nao.cas.cn} and B.F. Liu $^{1,2}$\\
$^{1}$Key Laboratory of Space Astronomy and Technology, National Astronomical Observatories, Chinese Academy of
Sciences, Beijing 100012, China \\
$^{2}$School of Astronomy and
Space Sciences, University of Chinese Academy of Sciences, 19A Yuquan Road, Beijing 100049, China\\} 

\date{Accepted XXX. Received YYY; in original form ZZZ}
\pubyear{2020}

\begin{document}
\label{firstpage}
\pagerange{\pageref{firstpage}--\pageref{lastpage}}
\maketitle
\begin{abstract}
Observationally, an anti-correlation between the X-ray photon index $\Gamma$ (obtained by fitting the 
X-ray spectrum between 0.5 and 10 keV with a single power law) and the X-ray luminosity $L_{\rm 0.5-10keV}$, 
i.e., a softening of the X-ray spectrum with decreasing $L_{\rm 0.5-10keV}$, is found in neutron star 
low-mass X-ray binaries (NS-LMXBs) in the range of $L_{\rm 0.5-10keV}\sim 10^{34}-10^{36}\ \rm erg\ s^{-1}$.
In this paper, we explain the observed anti-correlation between $\Gamma$ and $L_{\rm 0.5-10keV}$ 
within the framework of the self-similar solution of the advection-dominated accretion flow (ADAF) 
around a weakly magnetized NS. The ADAF model intrinsically predicts an anti-correlation between 
$\Gamma$ and $L_{\rm 0.5-10keV}$. 
In the ADAF model, there is a key parameter, $f_{\rm th}$, which describes the fraction of the ADAF 
energy released at the surface of the NS as thermal emission to be scattered in the ADAF. 
We test the effect of $f_{\rm th}$ on the anti-correlation between $\Gamma$ and $L_{\rm 0.5-10keV}$. 
It is found that the value of $f_{\rm th}$ can significantly affect the anti-correlation between 
$\Gamma$ and $L_{\rm 0.5-10keV}$. Specifically, the anti-correlation between $\Gamma$ and $L_{\rm 0.5-10keV}$ 
becomes flatter with decreasing $f_{\rm th}$ as taking $f_{\rm th}=0.1, 0.03, 0.01, 0.005$, $0.003$ and $0$ 
respectively.
By comparing with a sample of non-pulsating NS-LMXBs with well measured  
$\Gamma$ and $L_{\rm 0.5-10keV}$, we find that indeed only a small value of $0.003\lesssim f_{\rm th}\lesssim 0.1$ 
is needed to match the observed anti-correlation between $\Gamma$ and $L_{\rm 0.5-10keV}$.
Finally, we argue that the small value of $f_{\rm th}\lesssim 0.1$ derived in this paper further confirms 
our previous conclusion that the radiative efficiency of NSs with an ADAF accretion may not be as high 
as $\epsilon \sim {\dot M GM\over R_{*}}/{\dot M c^2}\sim 0.2$. 
\end{abstract}



\section{Introduction}
Neutron star low-mass X-ray binaries (NS-LMXBs) are binaries in which a NS
accretes matter from its low-mass companion star ($\lesssim 1 M_{\odot}$) via the Roche lobe. Most 
NS-LMXBs are X-ray transients, spending most of their time in the quiescent state
(often taken as the X-ray luminosity $\lesssim 10^{34}\ \rm erg \ s^{-1}$, and the X-ray luminosity 
here referring to the luminosity between 0.5 and 10 keV) for years to decades. 
NS-LMXBs are generally discovered as they go into outburst, during which the X-ray luminosity of 
some NS-LMXBs can generally increase by several orders of magnitude reaching up to 
$\sim 10^{36}-10^{38}\ \rm erg \ s^{-1}$. The outburst lasts from weeks to months,
and then the sources decay into quiescent state \citep[e.g.][for summary]{Campana1998,Wijnands2006}.
The X-ray spectrum of NS-LMXBs with the X-ray luminosity above $\sim 10^{36}\ \rm erg \ s^{-1}$
is generally well studied \citep[e.g][for discussions]{Lin2007}. While for NS-LMXBs in the low-luminosity  
regime as the X-ray luminosity below $\sim 10^{36}\ \rm erg \ s^{-1}$, due to the limitation of the
sensitivity of the X-ray instruments in orbit currently, the X-ray spectrum is relatively less well 
understood \citep[e.g.][]{Degenaar2013,ArmasPadilla2013b,DAngelo2015,Chakrabarty2014,Wijnands2015}.  
In this paper, we focus on the X-ray spectrum of NS-LMXBs between the quiescent state 
and the outburst in the X-ray luminosity range of $\sim 10^{34}-10^{36}\ \rm erg \ s^{-1}$.

Generally, the X-ray spectrum between 0.5 and 10 keV of NS-LMXBs in the X-ray luminosity range of
$L_{\rm 0.5-10\rm keV} \sim 10^{34}-10^{36}\ \rm erg \ s^{-1}$ can be satisfactorily described by a single 
power law, although the spectral fitting can be improved if an additional thermal soft X-ray component is 
added for the sources with sufficiently good data available,
especially in the range of $L_{\rm 0.5-10\rm keV} \sim 10^{34}-10^{35}\ \rm erg \ s^{-1}$ 
\citep[e.g.][]{ArmasPadilla2013b,Degenaar2013}. 
\citet[][]{Wijnands2015} compiled a sample of NS-LMXBs from literatures with well measured 
X-ray spectra between 0.5 and 10 keV in the range of 
$L_{\rm 0.5-10\rm keV} \sim 10^{34}-10^{36}\rm \ erg \ s^{-1}$.
All the sources in the sample are non-pulsating systems, which means that the effect of the magnetic field on the 
X-ray spectrum is very little. The X-ray spectra of all the sources in the sample are once reported in the 
literatures to be fitted with a single power-law model, i.e., $N(E)\propto E^{-\Gamma}$ 
(with $\Gamma$ being the X-ray photon index between 0.5 and 10 keV ). 
Then it is found that systematically there is a strong anti-correlation between the X-ray photon index $\Gamma$ 
and the X-ray luminosity $L_{\rm 0.5-10keV}$ for the sources in the sample as a whole in the range of 
$L_{\rm 0.5-10keV}\sim 10^{34}-10^{36}\ \rm erg \ s^{-1}$.

Such an anti-correlation between the X-ray photon index $\Gamma$ and the X-ray luminosity $L_{\rm X}$
\footnote{Please note that in different literatures, $\Gamma$ and $L_{\rm X}$ are sometimes measured 
in different energy bands, such as \citet[][]{Islam2018} for discussions.}
has also been widely reported for black hole (BH) X-ray transients in their hard state
\citep[e.g.][]{Wu2008,Plotkin2013,Homan2013,Liuhao2019}.
In \citet[][]{Wijnands2015}, the authors further compared the $\Gamma$-$L_{\rm 0.5-10keV}$
anti-correlation between NS-LMXBs and BH X-ray transients, from which it is found that NS-LMXBs have a significant 
softer X-ray spectrum for a fixed X-ray luminosity than that of BH X-ray transients. Additionally, the 
separation of the X-ray photon index $\Gamma$ between NS-LMXBs and BH X-ray transients 
is more obvious for the X-ray luminosity below $\sim 10^{35}\rm \ erg \ s^{-1}$ than that of 
above $\sim 10^{35}\rm \ erg \ s^{-1}$ (equivalent to that the slope of the $\Gamma$-$L_{\rm 0.5-10keV}$
anti-correlation in NS-LMXBs is steeper than that of in BH X-ray transients). 
The anti-correlation between $\Gamma$ and $L_{\rm X}$ (or $L_{\rm X}/L_{\rm Edd}$)
\footnote {$L_{\rm X}$ is often replaced by the Eddington scaled X-ray luminosity 
$L_{\rm X}/L_{\rm Edd}$ (with $L_{\rm Edd}$ being the Eddington luminosity, 
and $L_{\rm Edd}=1.26\times 10^{38} M/M_{\odot}\ \rm {erg\ s^{-1}}$), and
then the correlation between $\Gamma$ and $L_{\rm X}/L_{\rm Edd}$ can be  
extended up to supermassive BHs, as that of the anti-correlation between $\Gamma$ and $L_{\rm X}/L_{\rm Edd}$ 
established in low-luminosity active galactic nuclei (AGNs) \citep[e.g.][]{Gu2009,Younes2011,Veledina2011,
Emmanoulopoulos2012,Hernandez2013,Hernandez2014,Jang2014}.} 
in BHs has been well explained in the framework of the advection-dominated accretion flow (ADAF) 
\citep[e.g.][]{Qiao2010,Qiao2013,Qiaoetal2013,Yang2015}. ADAF is a kind of geometrically thick, 
optically thin, hot accretion flow \citep[][for review]{Ichimaru1977,Rees1982,Narayan1994,Narayan1995b,
Abramowicz1995,Manmoto1997,Yuan2014}, 
which is different from the geometrically thin, optically thick, cool accretion disc \citep[][]{Shakura1973}.  
For the ADAF around a BH, the electron temperature $T_{\rm e}$ of the ADAF increases very slightly with 
decreasing $\dot m$ ($\dot m=\dot M/\dot M_{\rm Edd}$, with
$\dot M_{\rm Edd}=L_{\rm Edd}/{0.1c^2}$=$1.39 \times 10^{18} M/M_{\rm \odot} \rm \ g \ s^{-1}$), 
while the Compton scattering optical depth $\tau_{\rm es}$ decreases with decreasing $\dot m$ by an equal multiple. 
So the Compton $y$-parameter (defined as $y={{4kT_{\rm e}}\over {m_{\rm e}c^2}} \tau_{\rm es}$, with $\tau_{\rm es}<1$)
decreases with decreasing $\dot m$, leading to a softer X-ray spectrum with decreasing $\dot m$. Meanwhile, in the 
ADAF solution $L_{\rm X}$ (or $L_{\rm {\rm X} }/L_{\rm Edd}$) decreases with decreasing $\dot m$, so a softer 
X-ray spectrum is predicted with decreasing $L_{\rm X}$ (or $L_{\rm {\rm X} }/L_{\rm Edd}$), i.e., an 
anti-correlation between $\Gamma$ and $L_{\rm X}$ (or $L_{\rm {\rm X} }/L_{\rm Edd}$) is predicted.  

The physical mechanism for the anti-correlation between $\Gamma$ and $L_{\rm 0.5-10keV}$ 
in NS-LMXBs as proposed in \citet[][]{Wijnands2015} is not very clear. The softer X-ray spectrum and the
steeper slope of the $\Gamma-L_{\rm 0.5-10keV}$ anti-correlation in NS-LMXBs compared with that of in BH X-ray 
transients are speculated to be very probably resulted by the effect of the hard surface of the NS.
In this paper, we focus on the anti-correlation between $\Gamma$ and $L_{\rm 0.5-10keV}$ 
in NS-LMXBs, and explain it within the framework of the self-similar solution of the ADAF around a 
weakly magnetized NS \citep[][]{Qiao2018b,Qiao2020}. In \citet[][]{Qiao2020}, the authors updated
the calculation of \citet[][]{Qiao2018b} with the effect of NS spin considered.
One of the key points in \citet[][]{Qiao2018b,Qiao2020} is that the radiative feedback between the hard 
surface of the NS and the ADAF is considered, as suggested by several other authors previously
\citep[e.g.][]{Narayan1995b}. Roughly speaking, the feedback (thermal emission) from the 
surface of the NS will cool the ADAF itself, which makes the electron temperature of the ADAF around NSs 
is lower than that of around BHs, consequently predicting a softer X-ray spectrum
(in the band greater than $\sim 2$ keV) in NSs compared with that of in BHs. Meanwhile, in general the 
existence of the thermal component in soft X-ray band makes the X-ray spectrum between 0.5 and 10 keV much
softer (depending on the temperature of the thermal component) 
if the X-ray spectrum between 0.5 and 10 keV is simply fitted with a single power law. 
Specifically, in \citet[][]{Qiao2020}, a fraction, $f_{\rm th}$, of the ADAF energy (including the  
internal energy, the radial kinetic energy and the rotational energy) transferred onto
the surface of the NS is assumed to be thermalized at the surface of the NS as thermal emission to be 
scattered in the ADAF itself. The authors self-consistently calculate the structure and the corresponding 
emergent spectrum of the ADAF by considering the radiative feedback between the NS and the ADAF. 
Physically, the value of $f_{\rm th}$ is uncertain, and as will show in this paper, the value of 
$f_{\rm th}$ can significantly affect the structure and the corresponding emergent spectrum of the ADAF, 
consequently affecting the $\Gamma-L_{\rm 0.5-10keV}$ anti-correlation.

In this paper, we test the value of $f_{\rm th}$ on the X-ray spectrum and the slope of the 
$\Gamma-L_{\rm 0.5-10keV}$ anti-correlation. It is found that for a fixed X-ray luminosity, a softer X-ray 
spectrum, i.e., a larger $\Gamma$, is predicted for taking a larger value of $f_{\rm th}$. 
Meanwhile, the $\Gamma-L_{\rm 0.5-10keV}$ anti-correlation becomes flatter with decreasing  
$f_{\rm th}$ as taking $f_{\rm th}=0.1, 0.03, 0.01, 0.005$, $0.003$ and $0$ respectively.
By comparing with a sample of non-pulsating NS-LMXBs with well measured $\Gamma$ (obtained
by fitting the observed X-ray spectrum between 0.5 and 10 keV with a single power law) in the range of 
$L_{\rm 0.5-10keV}\sim 10^{34}-10^{36}\ \rm erg\ s^{-1}$, we find that indeed only a small value of 
$0.003 \lesssim f_{\rm th}\lesssim 0.1$ is needed to match the observed anti-correlation between 
$\Gamma$ and $L_{\rm 0.5-10keV}$.
The data in the sample are searched from literatures, which are summarized at 
Table 1 of \citet[][]{Wijnands2015}, Appendix of \citet[][]{Parikh2017} and 
\citet[][]{Beri2019} respectively.
Theoretically, in our model, there are two components, i.e., a thermal component and a power-law 
component, both of which can contribute a fraction of the X-ray luminosity between 0.5 and 10 keV respectively.   
We conclude that in the range of $L_{\rm 0.5-10keV}\sim 10^{34}-10^{35}\ \rm erg\ s^{-1}$, the observed 
softening of the X-ray spectrum with decreasing $L_{\rm 0.5-10keV}$ is due to the increase of the 
fractional contribution of the thermal soft X-ray component, which is supported by both the theoretical 
and observational results that the fractional contribution of the power-law component 
$\eta$ ($\eta\equiv L^{\rm power\ law}_{\rm 0.5-10\rm keV}/L_{\rm 0.5-10\rm keV}$) decreases with 
decreasing $L_{\rm 0.5-10keV}$. While in the range of $L_{\rm 0.5-10keV}\sim 10^{35}-10^{36}\ \rm erg\ s^{-1}$, 
our explanation for the observed softening of the X-ray spectrum with decreasing $L_{\rm 0.5-10keV}$ 
is a little uncertain, which is probably due to a complex relation between the thermal soft X-ray component and 
the power-law component, or probably dominantly due to the softening of the power-law component itself.
Finally, we argue that the derived value of $f_{\rm th}\lesssim 0.1$ in this paper further 
confirms our previous conclusion that the radiative efficiency of NSs with an ADAF accretion may not be as high as 
$\epsilon \sim {\dot M GM\over R_{*}}/{\dot M c^2}\sim 0.2$ as proposed in \citet[][]{Qiao2020}. 
The model is briefly introduced in Section 2. The results are shown in Section 3. 
The discussions are in Section 4 and the conclusions are in Section 5.

\section{The model}
We considered the ADAF accretion around a weakly magnetized NS within the framework of 
the self-similar solution as in \citet[][]{Qiao2018b}, which was further updated in 
\citet[][]{Qiao2020} with the effect of NS spin included. 
As in \citet[][]{Qiao2020}, the structure of the ADAF can be calculated by specifying the NS mass
$m$ ($m=M/M_{\odot}$), the NS radius $R_{*}$, rotational frequency of the NS $\nu_{\rm NS}$ 
(describing NS spin), the mass accretion rate $\dot m$, 
the viscosity parameter $\alpha$ and the magnetic parameter $\beta$ (with magnetic pressure 
$p_{\rm m}={B^2/{8\pi}}=(1-\beta)p_{\rm tot}$, $p_{\rm tot}=p_{\rm gas}+p_{\rm m}$) 
for describing the micro physics of ADAF, and $f_{\rm th}$ describing the fraction of the total energy 
of the ADAF, $L_{*}$, (including the internal energy, the radial kinetic energy and the rotational energy) 
transferred onto the surface of the NS to be thermalized as thermal emission to be scattered in the ADAF 
itself. One can refer to \citet[][]{Qiao2020} for the detailed expression of $L_{*}$.
The effective temperature of the thermal emission at the surface of the NS can be expressed as,
\begin{equation}\label{equ:T}
\begin{array}{l}
T_{*}=\bigl({L_{*}f_{\rm th}\over {4\pi R_{*}^{2}\sigma}}\bigr)^{1/4}.
\end{array}
\end{equation} 
Here it is assumed that the radiation from the surface of the NS is isotropic. $\sigma$ is the
Stefan–Boltzmann constant. 
In this paper, we set the NS mass $m=1.4$, and the NS radius $R_{*}=10$ km.
Here we would like to mention that it is suggested that the NS radii $R_{*}$ are 
$\sim 10-12$ km for $m\sim 1.4-1.5$ \citep[][for review]{Degenaar2018b}. Since the  
uncertainty of $R_{*}$ is very little for taking $m=1.4$, we expect that our main conclusions in this paper 
will not be strongly influenced for simply taking $m=1.4$ and $R_{*}=10$ km respectively.
The value of the rotational frequency of the NS $\nu_{\rm NS}$ can affect the rotational energy of the ADAF 
transferred onto the surface of the NS, which is proportional to the difference between the rotational frequency 
of the NS and the rotational frequency of the ADAF at its inner boundary \citep[equation 2 of][]{Qiao2020}.
Since the angular velocity of the ADAF is sub-Keplerian, intrinsically the rotational energy of the 
ADAF transferred onto the surface of the NS is very small. Meanwhile, as shown in Section 4.1 of 
\citep[][]{Qiao2020}, the effect of $\nu_{\rm NS}$ on the structure of the ADAF can nearly be 
neglected as for taking $\nu_{\rm NS}=0, 200, 500$ and $700$ Hz respectively. In this paper, we simply 
set $\nu_{\rm NS}=0$.
The value of the viscosity parameter $\alpha$ is uncertain.  
In general, it is suggested that the value of the viscosity parameter is 
$\alpha\lesssim$ 1 \citep[][for discussions]{Shakura1973,Frank2002}. 
As we can see from \citet[][]{Qiao2020}, the upper limit of the X-ray luminosity $L_{\rm 0.5-10keV}$
predicted by the ADAF model is related with $\alpha$ [roughly with 0.1$\alpha^2$, 
see \citet[][]{Narayan1995b} for analysis]. In order to safely cover the upper limit of the X-ray 
luminosity of $L_{\rm 0.5-10keV}\sim 10^{36}\ \rm erg\ s^{-1}$ that we focus on in this paper. 
We simply take the suggested maximum value of the viscosity parameter, i.e., $\alpha=1$ 
throughout the paper.
In this paper, we set $\beta=0.95$ as suggested that the magnetic field in the ADAF solution is 
very weak by magnetohydrodynamic simulations \citep[][for review]{Yuan2014}.
So we have two parameters left, i.e., $f_{\rm th}$ and $\dot m$. 
Finally, we calculate the emergent spectrum of the model with the method of multiscattering of soft photons 
in the hot gas as in \citet[][]{Qiao2018b,Qiao2020}.

\section{Results}
\subsection{The effect of $f_{\rm th}$}\label{s:fth}
\begin{figure*}
\includegraphics[width=85mm,height=60mm,angle=0.0]{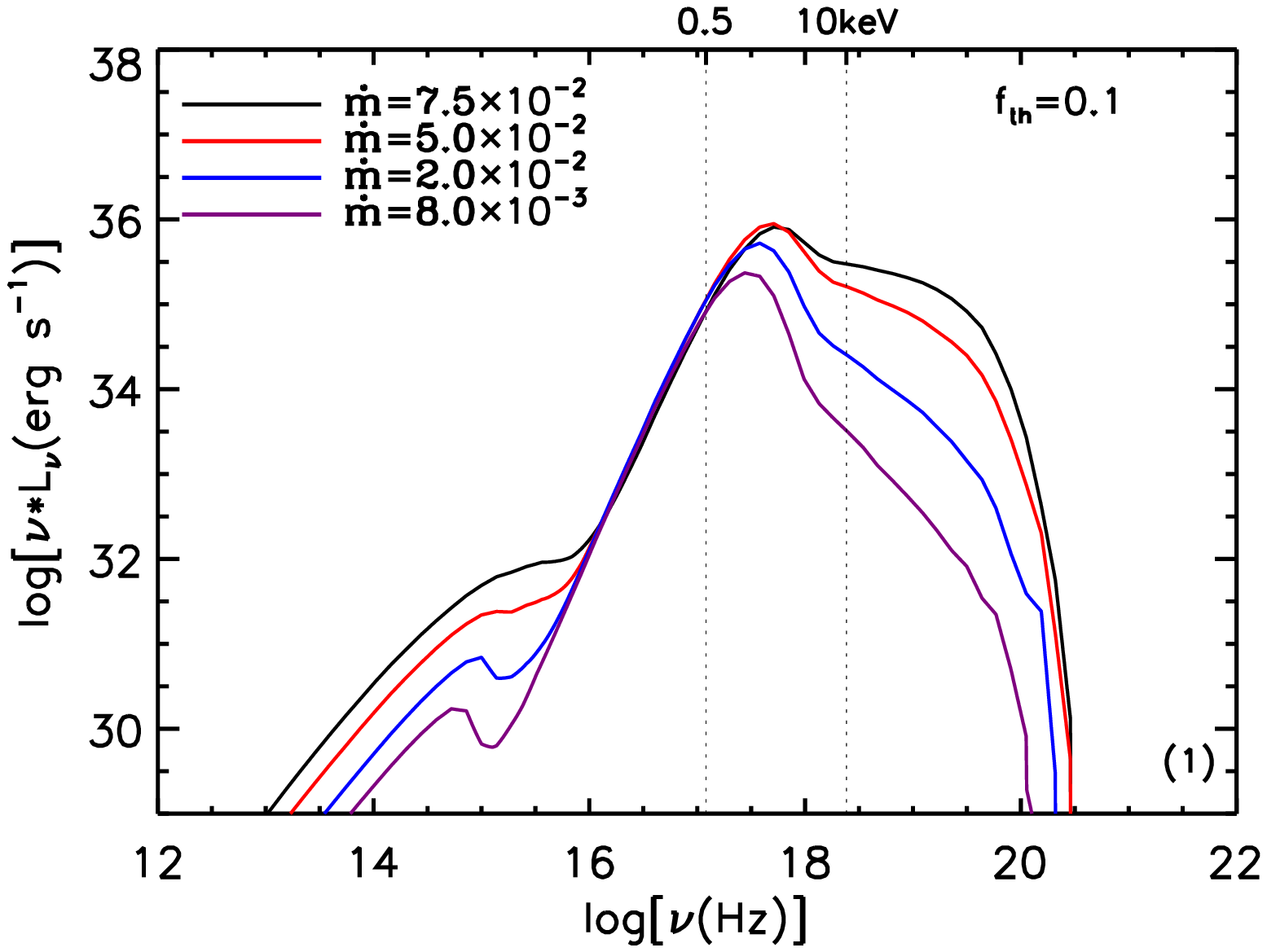}
\includegraphics[width=85mm,height=60mm,angle=0.0]{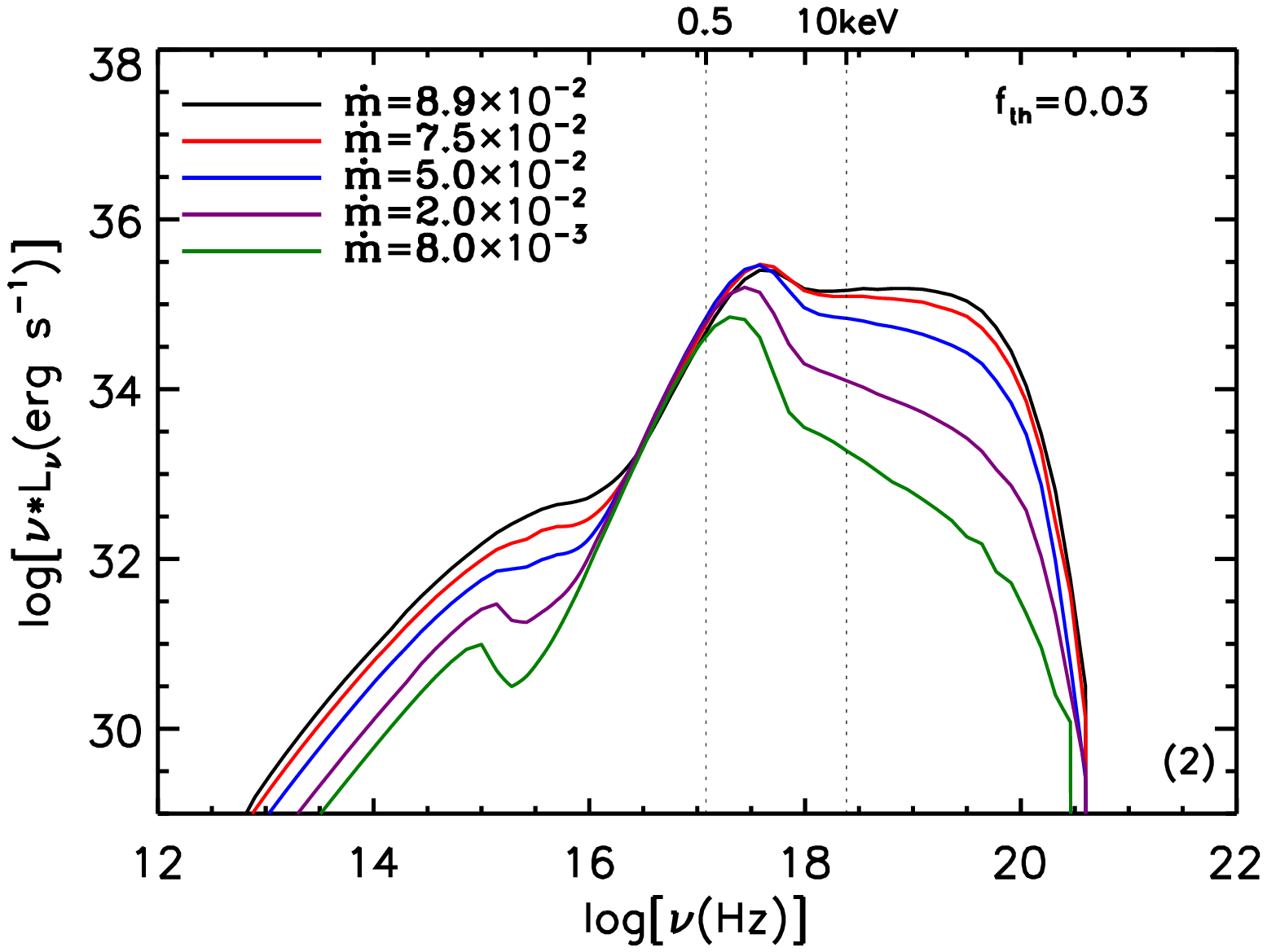}
\includegraphics[width=85mm,height=60mm,angle=0.0]{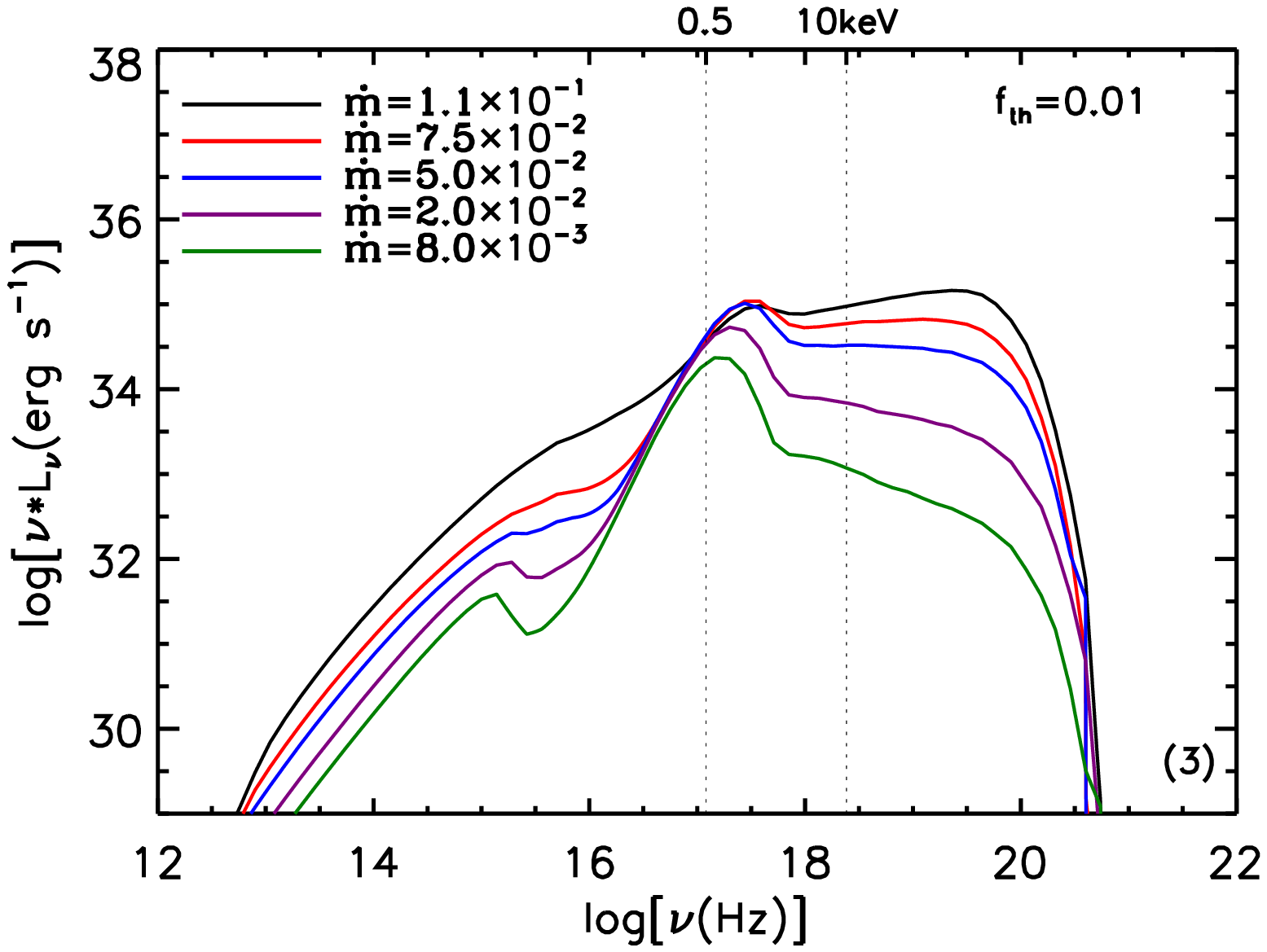}
\includegraphics[width=85mm,height=60mm,angle=0.0]{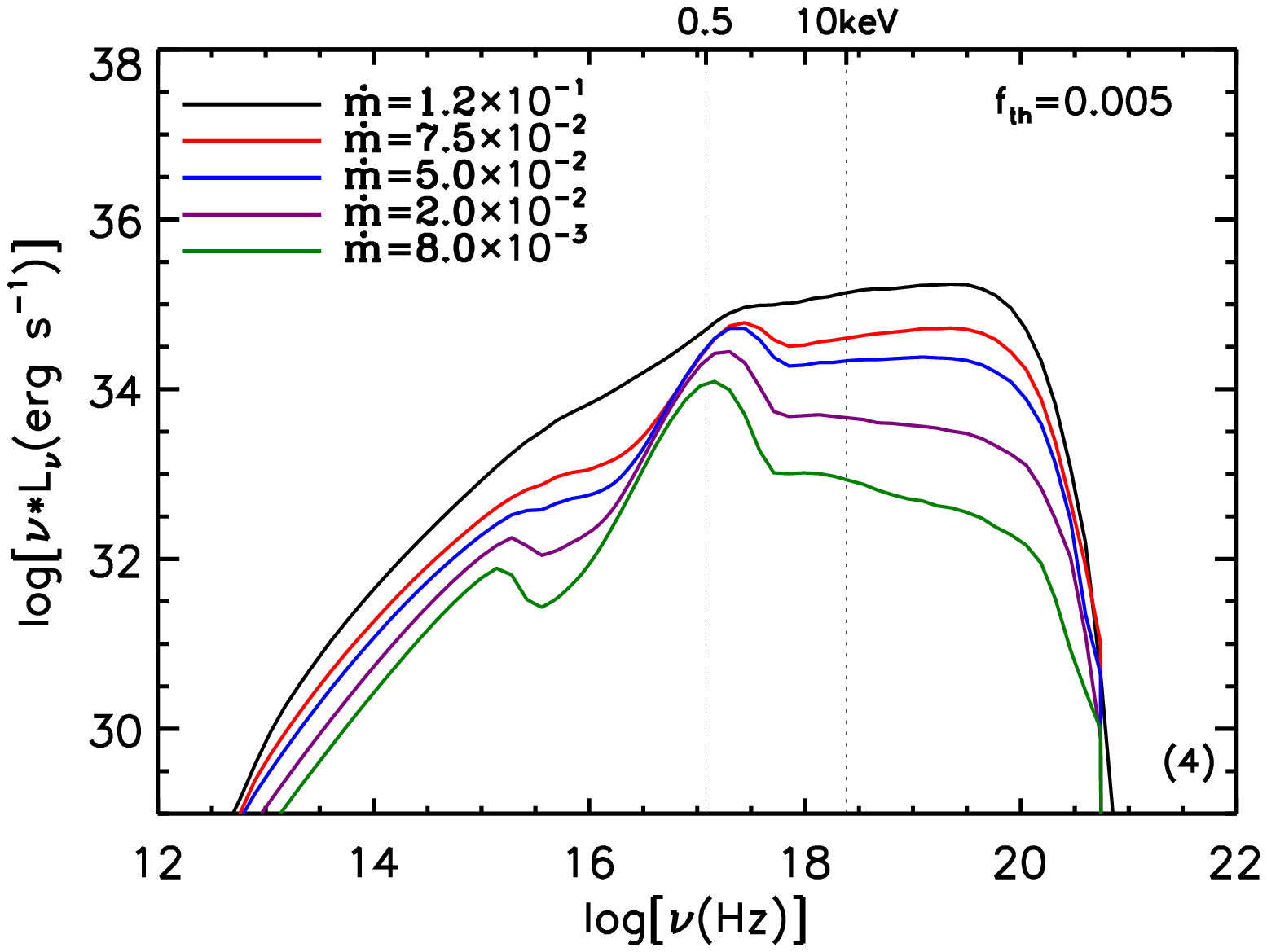}
\includegraphics[width=85mm,height=60mm,angle=0.0]{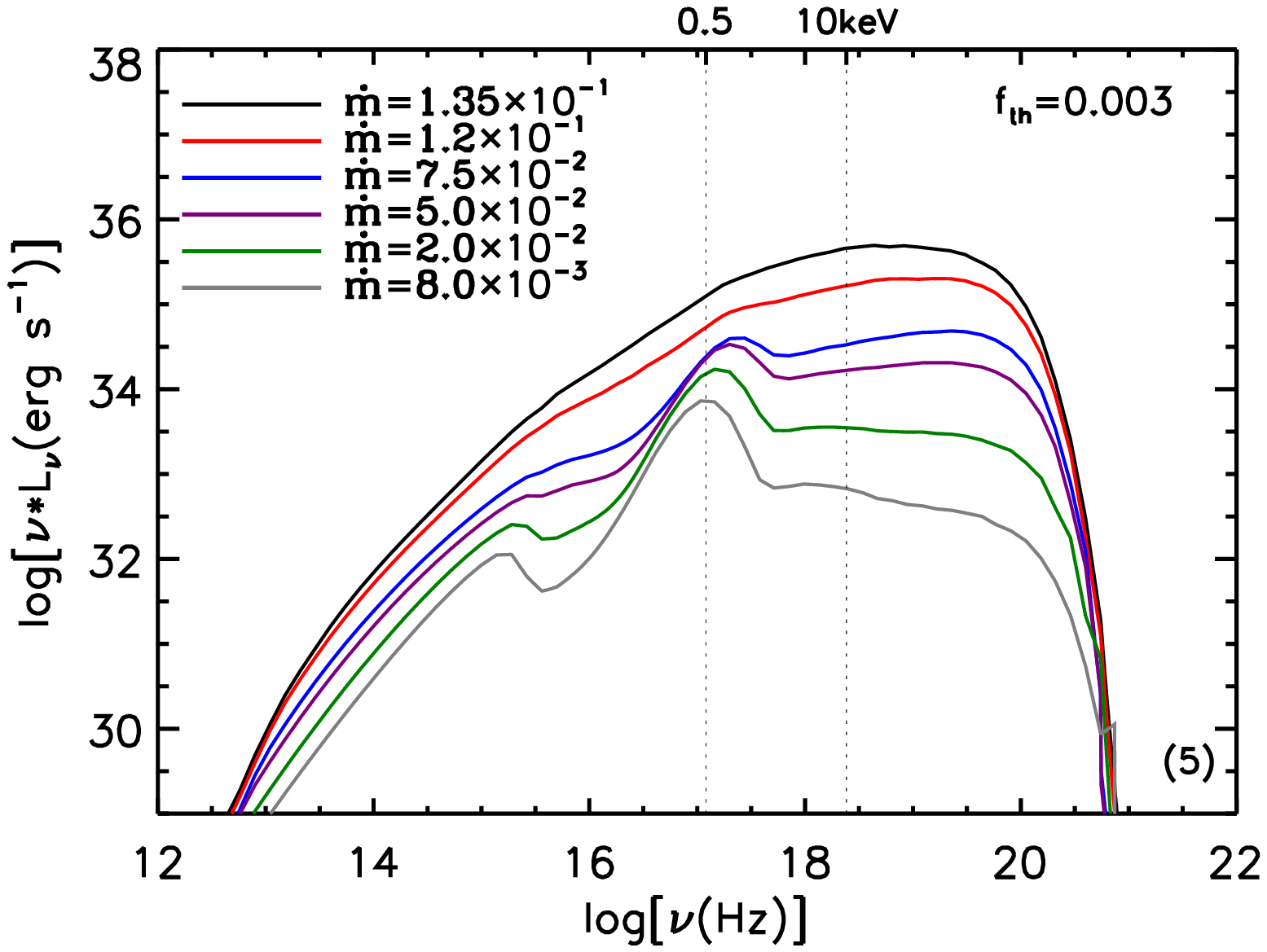}
\includegraphics[width=85mm,height=60mm,angle=0.0]{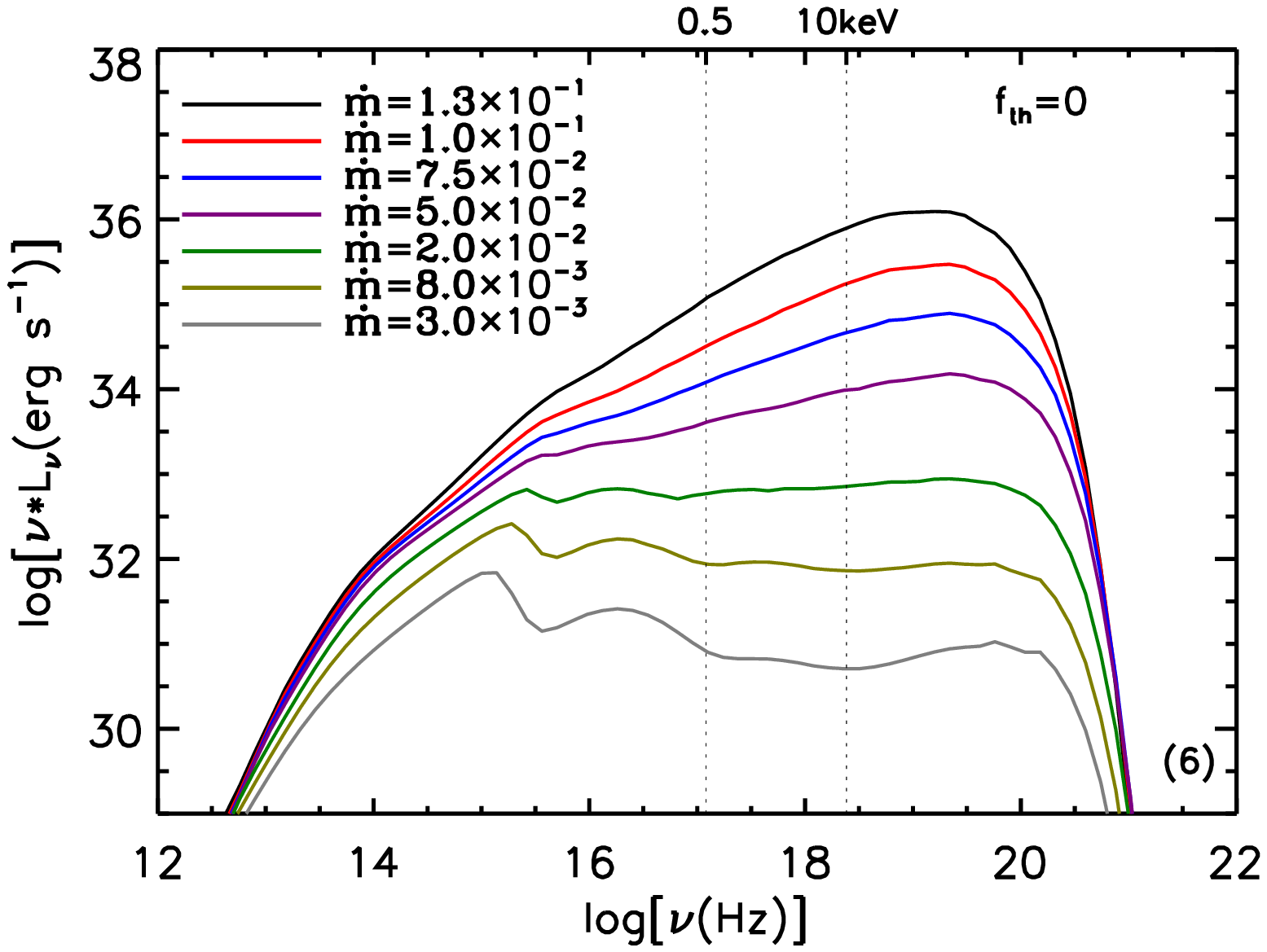}
\caption{\label{f:fth_sp}Panel (1): Emergent spectra of the ADAF around a weakly magnetized NS for $f_{\rm th}=0.1$.  
Panel (2): Emergent spectra of the ADAF around a weakly magnetized NS for $f_{\rm th}=0.03$.
Panel (3): Emergent spectra of the ADAF around a weakly magnetized NS for $f_{\rm th}=0.01$.
Panel (4): Emergent spectra of the ADAF around a weakly magnetized NS for $f_{\rm th}=0.005$.
Panel (5): Emergent spectra of the ADAF around a weakly magnetized NS for $f_{\rm th}=0.003$.
Panel (6): Emergent spectra of the ADAF around a weakly magnetized NS for $f_{\rm th}=0$. In all the calculations, 
we take $m=1.4$, $R_{*}=10\ \rm km$, $\alpha=1$ and $\beta=0.95$.}
\end{figure*}

\begin{figure}
\includegraphics[width=85mm,height=60mm,angle=0.0]{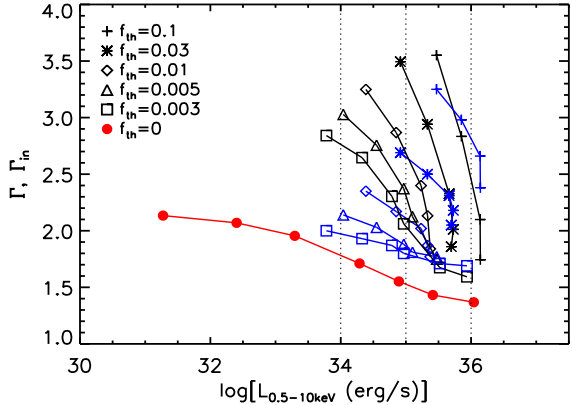}
\caption{\label{f:Gamma_L_theory}X-ray photon index $\Gamma$ (black symbol, obtained by fitting the 
X-ray spectrum of the model between 0.5 and 10 keV with a single power law) as a function of the X-ray luminosity 
$L_{\rm 0.5-10keV}$, and photon index of the intrinsic power-law component $\Gamma_{\rm in}$ (blue symbol, obtained 
by analyzing the X-ray spectrum of the model between 0.5 and 10 keV with a thermal soft
X-ray component added to the power-law component) as a function of $L_{\rm 0.5-10keV}$. 
For $f_{\rm th}=0$, the value of $\Gamma$ is same with $\Gamma_{\rm in}$, 
see the red filled-circle for $\Gamma$ (or $\Gamma_{\rm in}$) as a function of 
$L_{\rm 0.5-10keV}$.
}
\end{figure}

\begin{figure}
\includegraphics[width=85mm,height=60mm,angle=0.0]{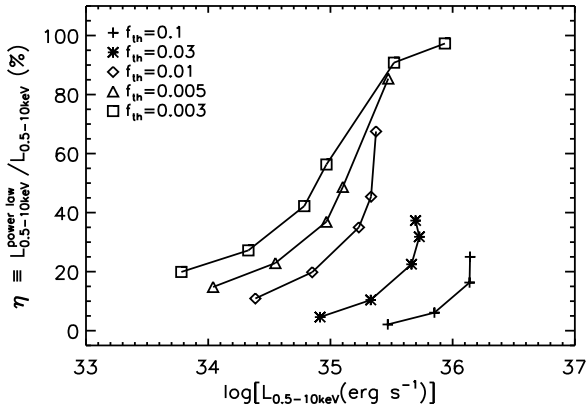}
\caption{\label{f:eta_L_theory}Fractional contribution of the power-law luminosity 
$\eta$ as a function of the X-ray luminosity $L_{\rm 0.5-10\rm keV}$. 
}
\end{figure}

In the panel (1) of Fig. \ref{f:fth_sp}, we plot the emergent spectra for $\dot m=7.5\times 10^{-2}$,
$5.0\times 10^{-2}$, $2.0\times 10^{-2}$ and $8.0\times 10^{-3}$ respectively with $f_{\rm th}=0.1$.
Specifically, from the emergent spectra, it is found that the X-ray photon index $\Gamma$ (obtained by fitting the X-ray 
spectrum between 0.5 and 10 keV with a single power law) increases from 1.74 to 3.56 with $L_{\rm 0.5-10keV}$ 
decreasing from $1.39\times 10^{36}$ to $2.95\times 10^{35}\ \rm erg\ s^{-1}$ 
(please note that throughout the paper $L_{\rm 0.5-10keV}$ is calculated by integrating the theoretical 
emergent spectrum of the ADAF model). 
One can refer to Table \ref{t:fth_effect} for the detailed numerical results of $\Gamma$. 
One can also refer to the black symbol `+' in Fig. \ref{f:Gamma_L_theory}
for $\Gamma$ versus $L_{\rm 0.5-10keV}$ for clarity. It is clear that 
there is an anti-correlation between $\Gamma$ and $L_{\rm 0.5-10keV}$.
However, intrinsically, from the emergent spectra of our model, there are two components, i.e., 
a thermal soft X-ray component and a power-law component, both of which can contribute to the X-ray 
luminosity between 0.5 and 10 keV.  It can be seen that the effective temperature $T_{*}$ of the 
thermal emission at the surface of the NS decreases from 
$0.62$ to $0.36$ keV with $L_{\rm 0.5-10keV}$ decreasing from $1.39\times 10^{36}$ to 
$2.95\times 10^{35}\ \rm erg\ s^{-1}$. 
One can refer to Table \ref{t:fth_effect} for the detailed numerical results of $T_{*}$.
Meanwhile, we show that the fractional contribution of the 
power-law component $\eta$ ($\eta\equiv L^{\rm power\ law}_{\rm 0.5-10\rm keV}/L_{\rm 0.5-10\rm keV}$) decreases 
from $25.0\%$ to to $2.10\%$ with $L_{\rm 0.5-10keV}$ decreasing from $1.39\times 10^{36}$ to 
$2.95\times 10^{35}\ \rm erg\ s^{-1}$.
One can refer to Table \ref{t:fth_effect} for the detailed numerical results of $\eta$, and the 
symbol `+' in Fig. \ref{f:eta_L_theory} for $\eta$ versus $L_{\rm 0.5-10keV}$ for clarity. 
The decrease of $\eta$ with decreasing $L_{\rm 0.5-10keV}$ can be understood as follows. 
Actually, both the thermal soft X-ray luminosity and power-law luminosity decrease with decreasing 
$\dot m$, however, the decrease of the power-law luminosity is quicker than that of the thermal soft 
X-ray luminosity, resulting in a decrease of $\eta$ with decreasing $\dot m$.
Since $L_{\rm 0.5-10keV}$ decreases with decreasing $\dot m$, $\eta$ 
decreases with decreasing $L_{\rm 0.5-10keV}$.
We further show that the photon index of the intrinsic power-law component $\Gamma_{\rm in}$ (obtained by 
analyzing the X-ray spectrum of the model between 0.5 and 10 keV with a thermal soft
X-ray component added to the power-law component) increases from 2.38 to 3.25 with $L_{\rm 0.5-10keV}$ decreasing from 
$1.39\times 10^{36}$ to $2.95\times 10^{35}\ \rm erg\ s^{-1}$. One can refer to Table \ref{t:fth_effect} for 
the detailed numerical results of $\Gamma_{\rm in}$, and the blue symbol `+' in Fig. \ref{f:Gamma_L_theory} for 
$\Gamma_{\rm in}$ versus $L_{\rm 0.5-10keV}$ for clarity.

Similar calculations for the emergent spectra for different $\dot m$ with  
$f_{\rm th}=0.03, 0.01, 0.005, 0.003$ and $0$ are presented in the panel (2), (3), (4), (5) and (6) 
of Fig. \ref{f:fth_sp} respectively. One can refer to Table \ref{t:fth_effect} for the detailed numerical results
of $\Gamma$, $T_{*}$, $\eta$, $\Gamma_{\rm in}$ and $L_{\rm 0.5-10keV}$ for different $\dot m$ with 
$f_{\rm th}=0.03, 0.01, 0.005, 0.003$ and $0$ respectively \footnote{
Since systematically the radiative efficiency of the ADAF will increase with increasing $f_{\rm th}$, which
leads to the ADAF more easily to be collapsed, consequently resulting in the upper limit of $\dot m$ 
decrease with increasing $f_{\rm th}$. In this paper, we roughly take 
$\dot m=7.5\times 10^{-2}$ for $f_{\rm th}=0.1$,
$\dot m=8.9\times 10^{-2}$ for $f_{\rm th}=0.03$,
$\dot m=1.1\times 10^{-1}$ for $f_{\rm th}=0.01$,
$\dot m=1.2\times 10^{-1}$ for $f_{\rm th}=0.005$,
and $\dot m=1.35\times 10^{-1}$ for $f_{\rm th}=0.003$ respectively as the upper limits of $\dot m$. 
We take $\dot m=1.3\times 10^{-1}$ for $f_{\rm th}=0$ as the upper limit so that 
the predicted upper X-ray luminosity is roughly at $\sim 10^{36}\ \rm erg\ s^{-1}$ matching the upper
X-ray luminosity that we focus on in this paper.}. 
One can also refer to Fig. \ref{f:Gamma_L_theory} for $\Gamma$ (black symbol) and $\Gamma_{\rm in}$
(blue symbol) versus $L_{\rm 0.5-10keV}$, and Fig. \ref{f:eta_L_theory} for 
$\eta$ versus $L_{\rm 0.5-10keV}$ respectively for clarity.

As we can see from Fig. \ref{f:Gamma_L_theory}, it is very clear that there is 
an anti-correlation between $\Gamma$ and $L_{\rm 0.5-10keV}$ for $f_{\rm th}=0.1, 0.03, 0.01, 0.005$, $0.003$ and $0$ 
respectively. Meanwhile, the slope of the $\Gamma-L_{\rm 0.5-10keV}$ anti-correlation becomes flatter with 
decreasing $f_{\rm th}$. 
From Fig. \ref{f:Gamma_L_theory}, we can see that $\Gamma_{\rm in}$ increases with 
decreasing $L_{\rm 0.5-10keV}$ for different $f_{\rm th}$, meaning that the intrinsic power-law component softens 
with decreasing the X-ray luminosity for different $f_{\rm th}$. Meanwhile, it can be seen that the separation 
between $\Gamma$ and $\Gamma_{\rm in}$ becomes more and more clear with decreasing $L_{\rm 0.5-10keV}$
for different $f_{\rm th}$, suggesting that the softening of $\Gamma$ with decreasing $L_{\rm 0.5-10keV}$ is due 
to the increase of the fraction contribution of the thermal soft X-ray component 
for different $f_{\rm th}$, which is confirmed by the trend of $\eta$ versus $L_{\rm 0.5-10keV}$ as that 
$\eta$ decreases with decreasing $L_{\rm 0.5-10keV}$ for $f_{\rm th}=0.1, 0.03, 0.01, 0.005$ and $0.003$  
respectively as can be seen in Fig. \ref{f:eta_L_theory}. 

\begin{table*}
\caption{Radiative features of the ADAF around a weakly magnetized NS for different $\dot m$ with 
$f_{\rm th}=0.1, 0.03, 0.01, 0.005, 0.003$ and $0$ respectively.  
$\Gamma$ is the X-ray photon index between 0.5 and 10 keV obtained by fitting the X-ray spectrum with a 
single power law. $T_{*}$ is the effective temperature at the surface of the NS. 
$\eta\equiv L^{\rm power\ law}_{\rm 0.5-10\rm keV}/L_{\rm 0.5-10\rm keV}$ is the fractional contribution of 
the power-law luminosity. $\Gamma_{\rm in}$ is the photon index of the intrinsic power-law component obtained by 
analyzing the X-ray spectrum between 0.5 and 10 keV with a thermal soft X-ray component added to the 
power-law component. $L_{\rm 0.5-10 keV}$ is the luminosity between 0.5 and 10 $\rm keV$.}
\centering
\begin{tabular}{ccccccc}
\hline\hline
\multicolumn{6}{l}{$m=1.4$, $R_{*}=10\ \rm km$, $\alpha=1$, $\beta=0.95$ and $\nu_{\rm NS}=0\ (\rm Hz)$} \\
\hline
$f_{\rm th}$ & $\dot m$  & $\Gamma$ & $T_{*} \ (\rm keV)$  &
 $\eta$(\%) & $\Gamma_{\rm in}$ &  $L_{\rm 0.5-10 keV}\ (\rm erg \ s^{-1})$   \\
\hline
0.1  &  $7.5\times10^{-2}$     & 1.74    &0.62    & 25.0  &2.38     & $1.39\times 10^{36}$        \\
0.1  &  $5.0\times10^{-2}$     & 2.10    &0.57    & 16.3  &2.66     & $1.38\times 10^{36}$        \\
0.1  &  $2.0\times10^{-2}$     & 2.83    &0.46    & 6.10  &2.98     & $7.09\times 10^{35}$        \\
0.1  &  $8.0\times10^{-3}$     & 3.56    &0.36    & 2.10  &3.25     & $2.95\times 10^{35}$        \\
\hline                                                           
0.03  &  $8.9\times10^{-2}$    & 1.86    &0.48    & 37.3  &2.05     & $4.97\times 10^{35}$        \\
0.03  &  $7.5\times10^{-2}$    & 2.01    &0.46    & 31.8  &2.18     & $5.34\times 10^{35}$        \\
0.03  &  $5.0\times10^{-2}$    & 2.33    &0.42    & 22.5  &2.31     & $4.60\times 10^{35}$        \\
0.03  &  $2.0\times10^{-2}$    & 2.94    &0.34    & 10.4  &2.50     & $2.14\times 10^{35}$        \\
0.03  &  $8.0\times10^{-3}$    & 3.49    &0.27    & 4.60  &2.69     & $8.23\times 10^{34}$        \\
\hline                                                           
0.01  &  $1.1\times10^{-1}$    & 1.84    &0.38    & 67.5  &1.77     & $2.36\times 10^{35}$        \\
0.01  &  $7.5\times10^{-2}$    & 2.13    &0.35    & 45.4  &1.87     & $2.15\times 10^{35}$       \\
0.01  &  $5.0\times10^{-2}$    & 2.40    &0.32    & 35.0  &2.02     & $1.71\times 10^{35}$        \\
0.01  &  $2.0\times10^{-2}$    & 2.87    &0.26    & 19.8  &2.17     & $7.10\times 10^{34}$        \\
0.01  &  $8.0\times10^{-3}$    & 3.25    &0.20    & 10.9  &2.35     & $2.43\times 10^{34}$        \\
\hline                                                           
0.005  &  $1.2\times10^{-1}$   & 1.75    &0.33    & 85.4  &1.77     & $2.97\times 10^{35}$        \\
0.005  &  $7.5\times10^{-2}$   & 2.12    &0.30    & 48.7  &1.81     & $1.26\times 10^{35}$        \\
0.005  &  $5.0\times10^{-2}$   & 2.37    &0.27    & 36.9  &1.88     & $9.27\times 10^{34}$        \\
0.005  &  $2.0\times10^{-2}$   & 2.76    &0.22    & 22.9  &2.03     & $3.55\times 10^{34}$        \\
0.005  &  $8.0\times10^{-3}$   & 3.03    &0.17    & 14.8  &2.14     & $1.10\times 10^{34}$        \\
\hline                                                           
0.003  &  $1.35\times10^{-1}$  & 1.59    &0.30    & 97.3  &1.69     & $8.65\times 10^{35}$        \\
0.003  &  $1.2\times10^{-1}$   & 1.67    &0.29    & 90.8  &1.71     & $3.31\times 10^{35}$        \\
0.003  &  $7.5\times10^{-2}$   & 2.06    &0.26    & 56.3  &1.80     & $9.28\times 10^{34}$        \\
0.003  &  $5.0\times10^{-2}$   & 2.30    &0.24    & 42.2  &1.87     & $6.11\times 10^{34}$        \\
0.003  &  $2.0\times10^{-2}$   & 2.65    &0.19    & 27.2  &1.93     & $2.13\times 10^{34}$        \\
0.003  &  $8.0\times10^{-3}$   & 2.84    &0.15    & 19.9  &2.00     & $6.03\times 10^{33}$        \\
\hline                                                           
0      &  $1.3\times10^{-1}$   & 1.37    & -      & 100.0 &1.37    & $1.10\times 10^{36}$          \\
0      &  $1.0\times10^{-1}$   & 1.43    & -      & 100.0 &1.43    & $2.59\times 10^{35}$          \\
0      &  $7.5\times10^{-2}$   & 1.55    & -      & 100.0 &1.55    & $7.80\times 10^{34}$          \\
0      &  $5.0\times10^{-2}$   & 1.71    & -      & 100.0 &1.71    & $1.94\times 10^{34}$          \\
0      &  $2.0\times10^{-2}$   & 1.96    & -      & 100.0 &1.96    & $1.97\times 10^{33}$          \\
0      &  $8.0\times10^{-3}$   & 2.07    & -      & 100.0 &2.07    & $2.53\times 10^{32}$          \\
0      &  $3.0\times10^{-3}$   & 2.13    & -      & 100.0 &2.13    & $1.88\times 10^{31}$          \\
\hline\hline
\end{tabular}
\\
\label{t:fth_effect}
\end{table*}

\subsection{Comparison with observations for the anti-correlation between $\Gamma$ and
$L_{\rm 0.5-10\rm keV}$---a constraint to $f_{\rm th}$ }\label{s:observation}
As has been shown in Section \ref{s:fth}, we test the effect of $f_{\rm th}$ on the X-ray  
spectra and the anti-correlation between $\Gamma$ and $L_{\rm 0.5-10\rm keV}$ within the 
framework of the self-similar solution of the ADAF around a weakly magnetized NS.
It is found that, theoretically, the effect of $f_{\rm th}$ on the $\Gamma-L_{\rm 0.5-10\rm keV}$ 
anti-correlation is significant. In the following, we compare the theoretical results for the  
$\Gamma-L_{\rm 0.5-10\rm keV}$ anti-correlation with observations, further constraining the value of 
$f_{\rm th}$.

We collect a sample composed of fifteen non-pulsating NS-LMXBs from literatures with well measured X-ray 
spectra between 0.5 and 10 keV in the range of $L_{\rm 0.5-10keV}\sim 10^{34}-10^{36}\ \rm erg \ s^{-1}$.
Since only the non-pulsating NS-LMXBs are collected, we expect that the effects of the magnetic field on 
the X-ray spectra are very little. 
Eleven sources in the sample, i.e., AX J1754.2-2754,
1RXS J171824.2-402934,
1RXH J173523.7-354013,
XTE J1709-267,
IGR J17062-6143,
1RXS J170854.4-321857,
SAX J1753.5-2349,
Swift J174805.3-24463,
Aql X-1, 
IGR J17494-3030 
and XTE J1719-291 are from \citet[][]{Wijnands2015}.
Three sources, i.e., 1RXS J180408.9-342058, EXO 1745-248 and IGR J17361-4441 are from \citet[][]{Parikh2017}.
MAXI J1957+032 is from \citet[][]{Beri2019}. 
Twelve sources in the sample (except for IGR J17494-3030, XTE J1719-291 and  IGR J17361-4441 in the sample) 
are confirmed as NS binaries since the Type I X-ray bursts have been observed.
The nature of IGR J17494-3030, XTE J1719-291 and  IGR J17361-4441 are not confirmed yet, 
however the three sources are very likely to be NSs. One can refer to \citet[][]{ArmasPadilla2013c} for IGR J17494-3030, 
\citet[][]{ArmasPadilla2011} for XTE J1719-291, and \citet[][]{Wijnands2015} for IGR J17361-4441 for 
discussions in detail.

\begin{figure*}
\includegraphics[width=185mm,height=135mm,angle=0.0]{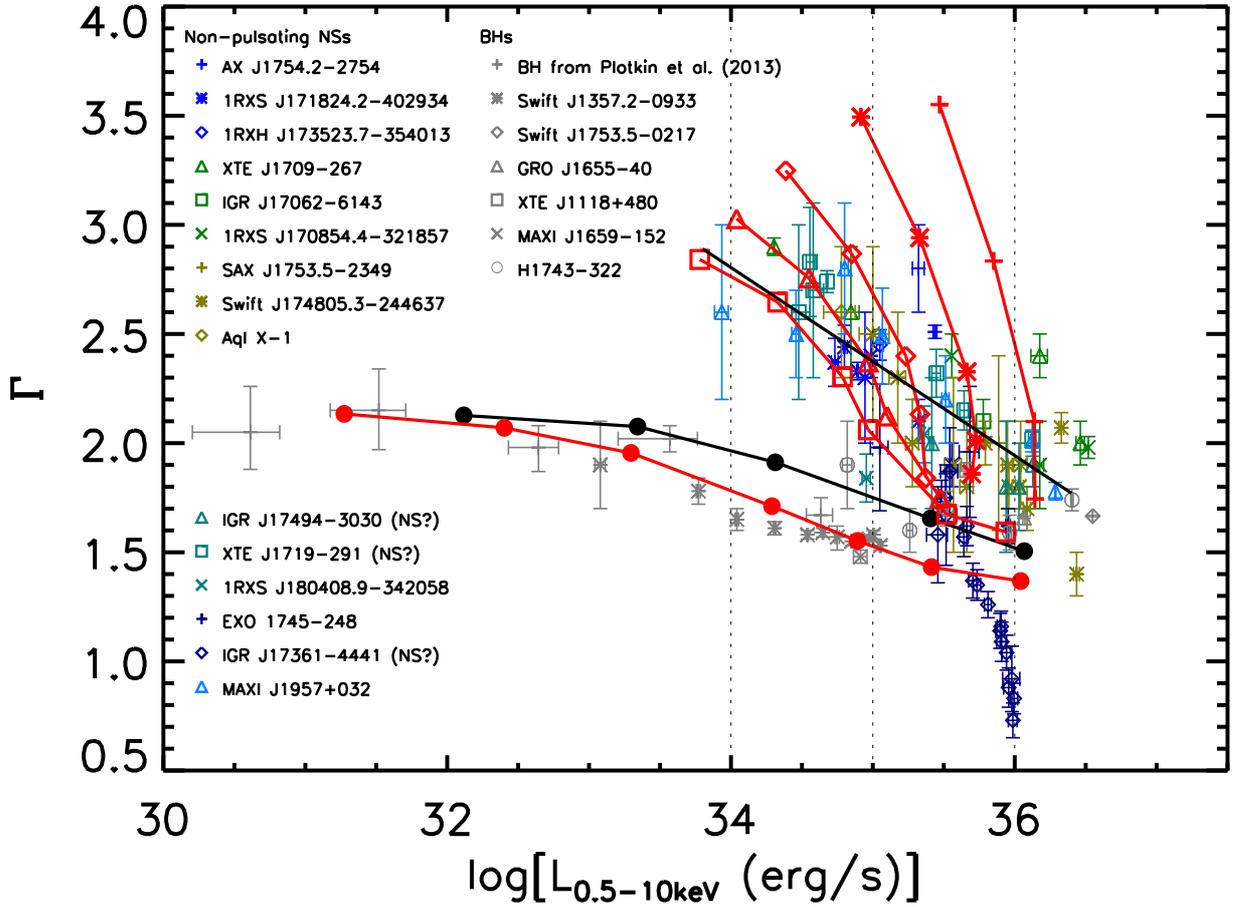}
\caption{\label{f:gama-L}X-ray photon index $\Gamma$ as a function of the X-ray 
luminosity $L_{\rm 0.5-10keV}$. The blue, dark green, dark yellow,  
dark cyan, dark blue and cyan blue symbols are the observational data for the fifteen 
non-pulsating NSs. The black solid line refers to the best-fitting linear regression of the observational 
data of the NSs (except for IGR J17361-4441). The red points of `+', `$\ast$', `$\diamond$', `$\triangle$', 
`$\square$' and the `red filled-circle' are the theoretical results of the ADAF around a weakly magnetized NS for 
$f_{\rm th}=0.1$, $0.03$, $0.01$, $0.005$ $0.003$ and $0$ respectively, and in the calculation 
$m=1.4$, $R_{*}=10\ \rm km$, $\alpha=1$ and $\beta=0.95$ are taken respectively. 
All the gray points are the observational data for BHs. The black filled-circle points are the   
theoretical results of the ADAF around a BH, and in the calculation 
$m=10$, $R_{\rm in}=3R_{\rm S}$, $\alpha=1$ and $\beta=0.95$ are taken respectively.  
}
\end{figure*}

The X-ray spectra between 0.5 and 10 keV of all the fifteen sources 
in the sample are once reported in the literatures to be fitted with a single power law, i.e., 
$N(E)\propto E^{-\Gamma}$ (with $N(E)$ being the
photon number at a given energy $E$ and $\Gamma$ being the X-ray photon index between 0.5 and 10 keV). 
In general, the spectral fitting of the $\it Swift$ X-ray data with a single power law can typically
result in an accepted fit. On the other hand, the X-ray spectra between 0.5 and 10 keV 
of seven sources in the sample, i.e., AX J1754.2-2754, 1RXS J171824.2-402934 and 1RXH J173523.7-354013
\citep[][]{ArmasPadilla2013b}, 
XTE J1709-267 \citep[][]{Degenaar2013},
Swift J174805.3-24463 \citep[][]{Bahramian2014},
IGR J17494-3030 \citep[][]{ArmasPadilla2013c},
and XTE J1719-291 \citep[][]{ArmasPadilla2011}
are once reported to be fitted with both the single power-law model and the two-component model, i.e., 
a thermal soft X-ray component at lower energies plus a power-law component at higher energies. 
Specifically, it is often found that if the high quality $\it XMM-Newton$ X-ray data can be available,  
the spectral fitting results can be significantly improved with a thermal soft X-ray component
added to the power-law component, especially in the range of 
$L_{\rm 0.5-10\rm keV} \sim 10^{34}-10^{35}\ \rm erg \ s^{-1}$.  
As for the relation of $\Gamma$ versus $L_{\rm 0.5-10\rm keV}$ we focus on in this paper,
we select the data for the X-ray spectra fitted with a single power law as in 
\citet[][]{Wijnands2015} and \citet[][]{Parikh2017}. 
The power-law X-ray photon index $\Gamma$ for all the data of the sources in the sample has an error less 
than 0.5 (which results in that several published sources with the measurements of $\Gamma$ are not 
included in this paper). 
One can refer to the blue, dark green, dark yellow, dark cyan, 
dark blue and cyan blue symbols in Fig. \ref{f:gama-L} for the observational data of 
$\Gamma$ versus $L_{\rm 0.5-10\rm keV}$ for details.
It is clear that there is an anti-correlation between $\Gamma$ and $L_{\rm 0.5-10\rm keV}$ 
(except for IGR J17361-4441). $\Gamma$ increases from $\sim 1.8$ to $\sim 3$ with $L_{\rm 0.5-10\rm keV}$ 
decreasing from $\sim 10^{36}\ \rm erg \ s^{-1}$ to $\sim 10^{34}\ \rm erg \ s^{-1}$. 
The best-fitting linear regression with the least square method
for the observational data between $\Gamma$ and $L_{\rm 0.5-10\rm keV}$ gives, 
\begin{eqnarray}\label{e:fit} 
\Gamma=-0.43\ \times\ {\rm log}\ L_{\rm 0.5-10 keV} + 17.4,
\end{eqnarray}
where the errors of the data are not considered. One can refer to the black solid line in 
Fig. \ref{f:gama-L} for clarity. One should also note that in fitting the data 
between $\Gamma$ and $L_{\rm 0.5-10\rm keV}$, we do not add IGR J17361-4441,
which may reflect a population of NSs with distinct very hard X-ray spectra, and will be 
discussed separately in Section \ref{s:hard}.  

We plot the theoretical results of $\Gamma$ versus $L_{\rm 0.5-10\rm keV}$ in Fig. \ref{f:gama-L} 
for comparisons. One can refer to the red points of `+', `$\ast$', `$\diamond$', `$\triangle$', 
`$\square$' and the `red filled-circle' for $f_{\rm th}=0.1$, $0.03$, $0.01$, $0.005$ $0.003$ and $0$ respectively. 
Roughly speaking, the observational data of $\Gamma$ versus $L_{\rm 0.5-10\rm keV}$ 
can be covered by the theoretical curves for taking $f_{\rm th}$ as 
$0.003\lesssim f_{\rm th}\lesssim 0.1$. 
Specifically, from Fig. \ref{f:gama-L}, it can be seen that 
in the range of $L_{\rm 0.5-10\rm keV}\sim 10^{35}-10^{36}\ \rm erg \ s^{-1}$, there are three crossing 
points between the theoretical curves of $\Gamma$ versus $L_{\rm 0.5-10\rm keV}$ and equation \ref{e:fit}, 
corresponding to the crossing point between equation \ref{e:fit} and the theoretical curve from right to 
left with $f_{\rm}=0.1$, $0.03$ and $0.01$ respectively. That is to say in the range of 
$L_{\rm 0.5-10\rm keV}\sim 10^{35}-10^{36}\ \rm erg \ s^{-1}$, $f_{\rm th}$ is roughly in the range of
$0.01\lesssim f_{\rm th}\lesssim 0.1$.
While in the range of $L_{\rm 0.5-10keV}\sim 10^{34}-10^{35}\ \rm erg \ s^{-1}$, 
although there is only one crossing point between equation \ref{e:fit} and the theoretical curve  
corresponding to $f_{\rm th}=0.005$, it can be seen that equation \ref{e:fit} can be matched for 
taking $f_{\rm th}$ as $0.003\lesssim f_{\rm th}\lesssim 0.005$. 

\begin{figure*}
\includegraphics[width=85mm,height=60mm,angle=0.0]{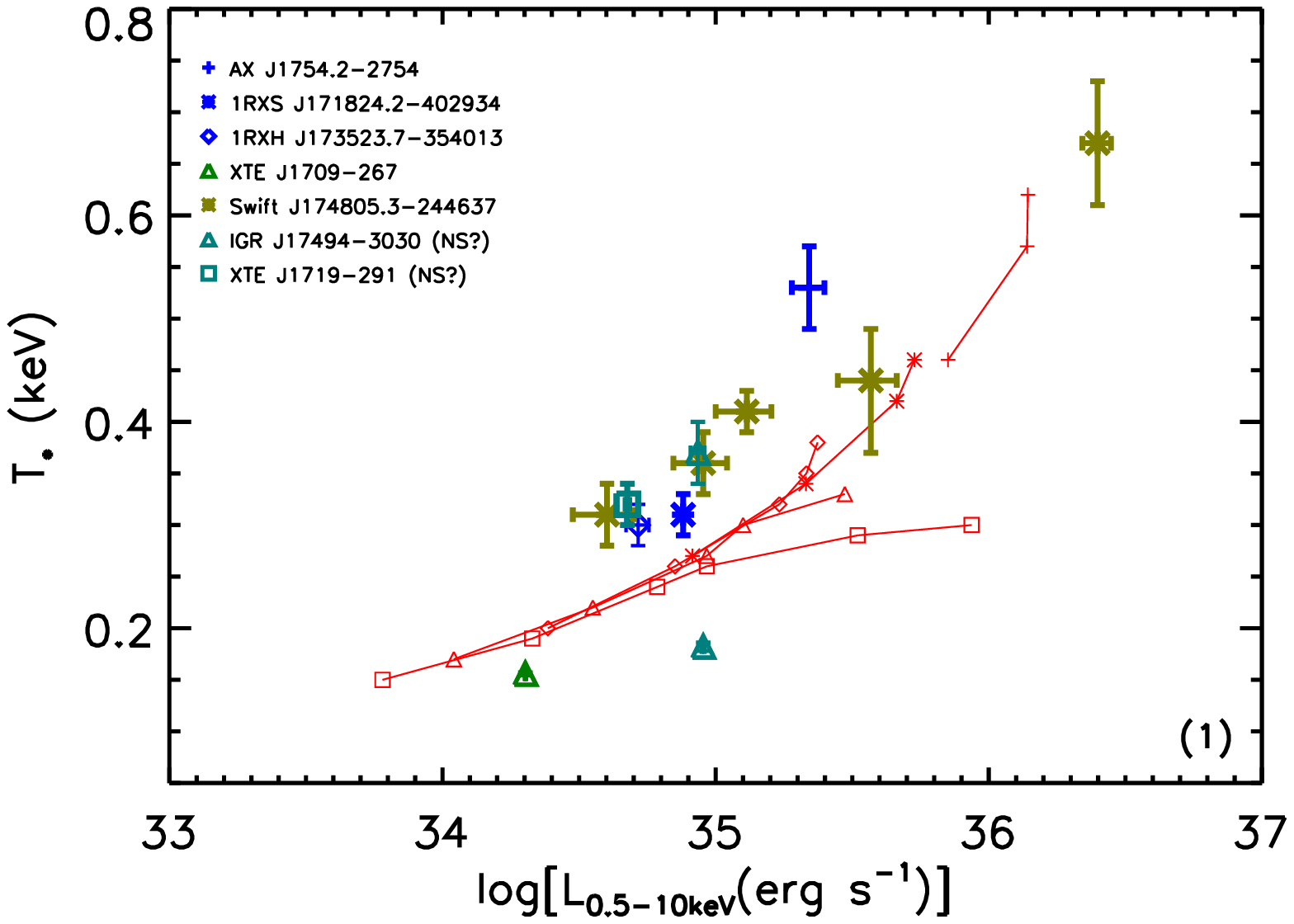}
\includegraphics[width=85mm,height=60mm,angle=0.0]{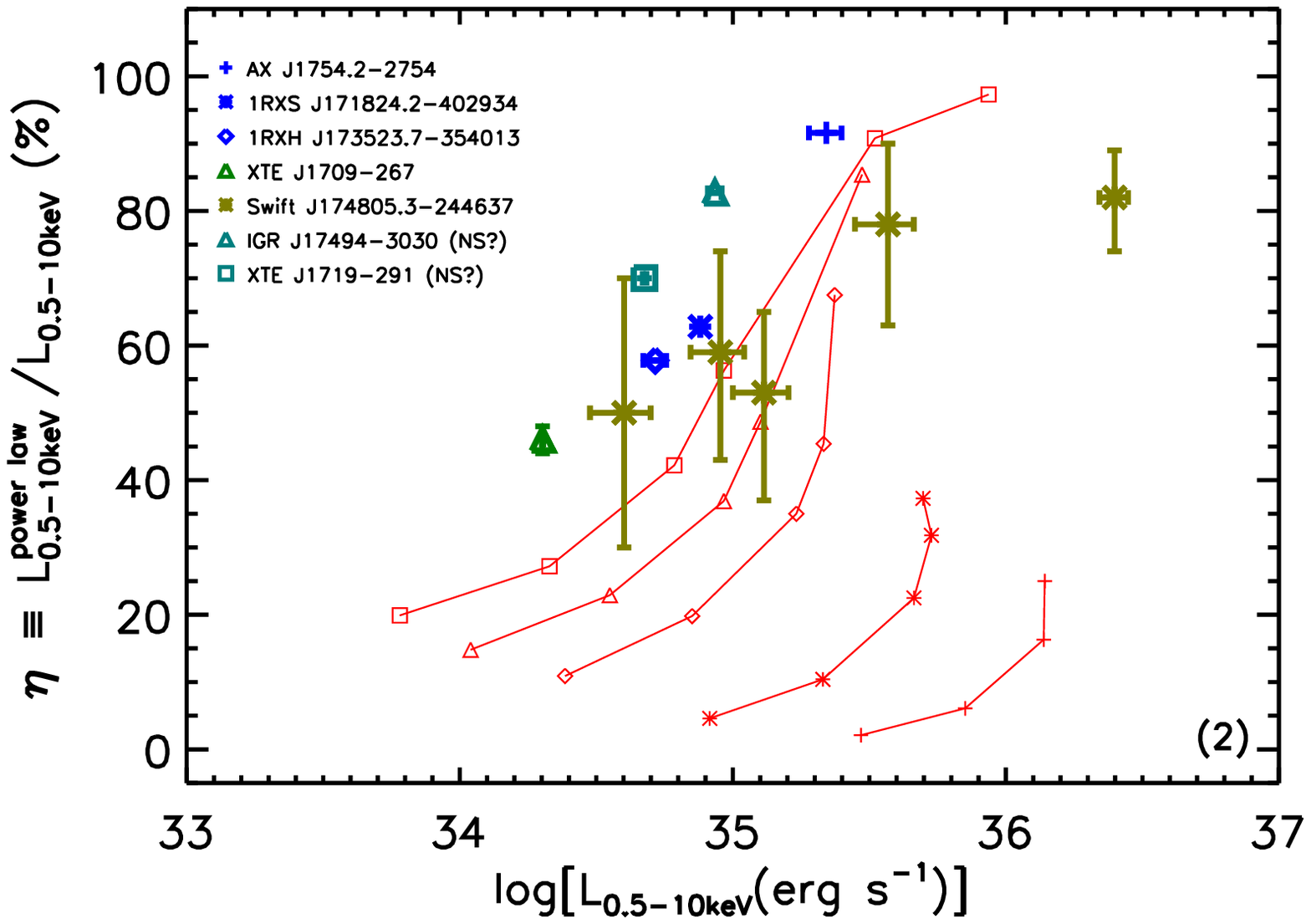}
\caption{\label{f:T-frac}Panel (1): Effective temperature at the surface of the NS $T_{*}$ 
as a function of the X-ray luminosity $L_{\rm 0.5-10keV}$. 
The blue, dark green, dark yellow and dark cyan symbols are the observational data.
The red points of `+', `$\ast$', `$\diamond$', `$\triangle$' and `$\square$' are the theoretical results 
of the ADAF around a weakly magnetized NS for $f_{\rm th}=0.1$, $0.03$, $0.01$, $0.005$ and $0.003$ respectively.
Panel (2): Fractional contribution of the power-law luminosity $\eta$ as a function of the X-ray 
luminosity $L_{\rm 0.5-10\rm keV}$. The blue, dark green, dark yellow and dark cyan symbols are the 
observational data. The red points of `+', `$\ast$', `$\diamond$', `$\triangle$' and 
`$\square$' are the theoretical results of the ADAF around a weakly magnetized NS for 
$f_{\rm th}=0.1$, $0.03$, $0.01$, $0.005$ and $0.003$ respectively.
}
\end{figure*}

In order to check the value of $f_{\rm th}$ constrained by the 
anti-correlation between $\Gamma$ and $L_{\rm 0.5-10keV}$ in the range of 
$L_{\rm 0.5-10keV}\sim 10^{34}-10^{36}\ \rm erg\ s^{-1}$, we further investigate the observational data 
of $T_{*}$ versus $L_{\rm 0.5-10keV}$ and $\eta$ versus 
$L_{\rm 0.5-10\rm keV}$ for comparison.
Specifically, we plot the observational data of  
$T_{*}$ versus $L_{\rm 0.5-10keV}$ in panel (1) of Fig. \ref{f:T-frac}, 
and $\eta$ versus $L_{\rm 0.5-10\rm keV}$ in panel (2) of Fig. \ref{f:T-frac} respectively
for the seven sources in the sample, i.e., AX J1754.2-2754, 1RXS J171824.2-402934, 1RXH J173523.7-354013, 
XTE J1709-267, Swift J174805.3-24463, IGR J17494-3030 and XTE J1719-291, the high quality X-ray spectra
of which are once reported to be fitted with a two-component model, i.e., a thermal soft X-ray component 
added to the power-law component.
The theoretical results of the ADAF model for $T_{*}$ versus $L_{\rm 0.5-10keV}$ and $\eta$ versus 
$L_{\rm 0.5-10\rm keV}$ are also plotted in panel (1) and panel (2) of Fig. \ref{f:T-frac} respectively 
as comparison. Specifically, the red points of `+', `$\ast$', `$\diamond$', `$\triangle$' and `$\square$' 
are the theoretical results for $f_{\rm th}=0.1$, $0.03$, $0.01$, $0.005$ and $0.003$ respectively. 

In general, it can be seen that in the range of $L_{\rm 0.5-10keV}\sim 10^{34}-10^{35}\ \rm erg\ s^{-1}$,  
despite the scatter, the value of $f_{\rm th}$ constrained by both the diagram of $T_{*}$ versus 
$L_{\rm 0.5-10keV}$ and $\eta$ versus $L_{\rm 0.5-10\rm keV}$ are roughly consistent with that of constrained 
by the anti-correlation between $\Gamma$ and $L_{\rm 0.5-10keV}$, i.e., $0.003\lesssim f_{\rm th}\lesssim 0.005$. 
As we can see from panel (2) of Fig. \ref{f:T-frac}, for $f_{\rm th}=0.003$, $\eta$
decreases from $\sim 60\%$ to $\sim 20\%$ for $L_{\rm 0.5-10keV}$ decreasing from $\sim 10^{35}\ \rm erg\ s^{-1}$
to $\sim 10^{34}\ \rm erg\ s^{-1}$, and for $f_{\rm th}=0.005$, $\eta$
decreases from $\sim 40\%$ to $\sim 15\%$ for $L_{\rm 0.5-10keV}$ decreasing from $\sim 10^{35}\ \rm erg\ s^{-1}$
to $\sim 10^{34}\ \rm erg\ s^{-1}$. 
That is to say, in general, $\eta$ decreases with decreasing  $L_{\rm 0.5-10keV}$ for $f_{\rm th}=0.003$ 
and $0.005$ respectively in the range of $L_{\rm 0.5-10keV}\sim 10^{34}-10^{35}\ \rm erg \ s^{-1}$.
Further, combining the result in Section \ref{s:fth} that $\Gamma_{\rm in}$ is 
smaller than $\Gamma$, and $\Gamma_{\rm in}$ as a function of $L_{\rm 0.5-10keV}$
is significantly deviated from $\Gamma$ as a function of $L_{\rm 0.5-10keV}$, 
we conclude that the softening of the X-ray spectrum is due to the increase of the fractional 
contribution of the thermal soft X-ray component in the range of 
$L_{\rm 0.5-10keV}\sim 10^{34}-10^{35}\ \rm erg\ s^{-1}$.  

While in the range of $L_{\rm 0.5-10keV}\sim 10^{35}-10^{36}\ \rm erg\ s^{-1}$, 
the case is a little complicated. 
In the diagram of $T_{*}$ versus $L_{\rm 0.5-10keV}$, the observed $T_{*}$ is slightly 
higher than that of the model predictions, for which we think it is very possible that we do not consider the 
thermal emission by the crust cooling of the NS itself in the model calculations. And further due to 
the change of $T_{*}$ is always in a very narrow range, we think $T_{*}$ is not good tracer for 
constraining the value of $f_{\rm th}$.
In the diagram of $\eta$ versus $L_{\rm 0.5-10\rm keV}$, as we can see from panel (2) of Fig. \ref{f:T-frac},
in the range of $L_{\rm 0.5-10keV}\sim 10^{35}-10^{36}\ \rm erg\ s^{-1}$, the observed value of $\eta$ is 
generally greater than $50\%$, meaning that in this luminosity range the X-ray spectrum is dominant by a power law. 
This requires a smaller value of $f_{\rm th} \lesssim 0.005$, 
which is inconsistent with the value of $0.01\lesssim f_{\rm th}\lesssim 0.1$ constrained by the 
anti-correlation between $\Gamma$ and $L_{\rm 0.5-10keV}$. Here we would like to mention that, 
based on our sample, only very few observational points fall in this luminosity range in 
the diagram of $\eta$ versus $L_{\rm 0.5-10\rm keV}$. We expect that more high quality $\it XMM-Newton$ 
X-ray data of the sources in our sample (even other sources) can be available to confirm such a 
discrepancy (or not) in the future. We would like to address further, if the constrained value of 
$f_{\rm th}$ is uncertain, the explanation for the softening of the X-ray spectrum is uncertain 
accordingly in the range of $L_{\rm 0.5-10keV}\sim 10^{35}-10^{36}\ \rm erg\ s^{-1}$. 
One can refer to Section \ref{s:discussion} for the detailed discussions.

As a comparison, the observational data of $\Gamma$ versus $L_{\rm 0.5-10\rm keV}$ for several BH X-ray 
transients are plotted in Fig. \ref{f:gama-L}. One can refer to the gray points in Fig. \ref{f:gama-L}. 
The data are summarized at Table 2 of \citet[][]{Wijnands2015}, including the BH sample
of \citet[][]{Plotkin2013}, as well as other six BH X-ray transients, i.e., 
Swift J1357.2-0933 \citep[][]{ArmasPadilla2013a},
Swift J1753.5-0217, Swift J1753.5-0217 and  GRO J1655-40 \citep[][]{Reis2010},
MAXI J1659-152 \citep[][]{Jonker2012} and H1743-322 \citep[][]{Jonker2010}. 
The error of $\Gamma$ for the BH sample is also less than 0.5 as that of the NS sample in this paper. 
The BH sample covers a wide X-ray luminosity range from a few times 
$10^{36}\ \rm erg \ s^{-1}$ down to a few times  $10^{30}\ \rm erg \ s^{-1}$. 
In general, there is an anti-correlation between $\Gamma$ and $L_{\rm 0.5-10\rm keV}$ 
in the range of $L_{\rm 0.5-10\rm keV} \sim 10^{34}-10^{36}\ \rm erg \ s^{-1}$. 
$\Gamma$ increases from $\sim 1.5$ to $\sim 2$ with $L_{\rm 0.5-10\rm keV}$ decreasing from 
$\sim 10^{36}\ \rm erg \ s^{-1}$ to $\sim 10^{34}\ \rm erg \ s^{-1}$. 
Below the X-ray luminosity of $L_{\rm 0.5-10\rm keV} \sim 10^{34}\ \rm erg \ s^{-1}$, the X-ray 
spectra gradually level off at an averaged value of $<\Gamma>\approx 2.1$.
In the X-ray luminosity range of 
$L_{\rm 0.5-10\rm keV} \sim 10^{34}-10^{36}\ \rm erg \ s^{-1}$ that we focus on in this paper,
it can be seen that, generally the X-ray spectra for the BH X-ray transients are harder 
than that of the NS-LMXBs. Especially, in the range of $L_{\rm 0.5-10\rm keV} \sim 10^{34}-10^{35}\ \rm erg \ s^{-1}$,
the separation between the BH X-ray transients and the NS-LMXBs is very clear, i.e., 
the X-ray spectra of the BH X-ray transients are much harder than that of the NS-LMXBs.
While in the range of $L_{\rm 0.5-10\rm keV} \sim 10^{35}-10^{36}\ \rm erg \ s^{-1}$, there is a significant 
overlap of the data points, and the separation between the BH X-ray transients and the NS-LMXBs is not 
as clear as that of in the range of $L_{\rm 0.5-10\rm keV} \sim 10^{34}-10^{35}\ \rm erg \ s^{-1}$. 

\begin{figure}
\includegraphics[width=85mm,height=60mm,angle=0.0]{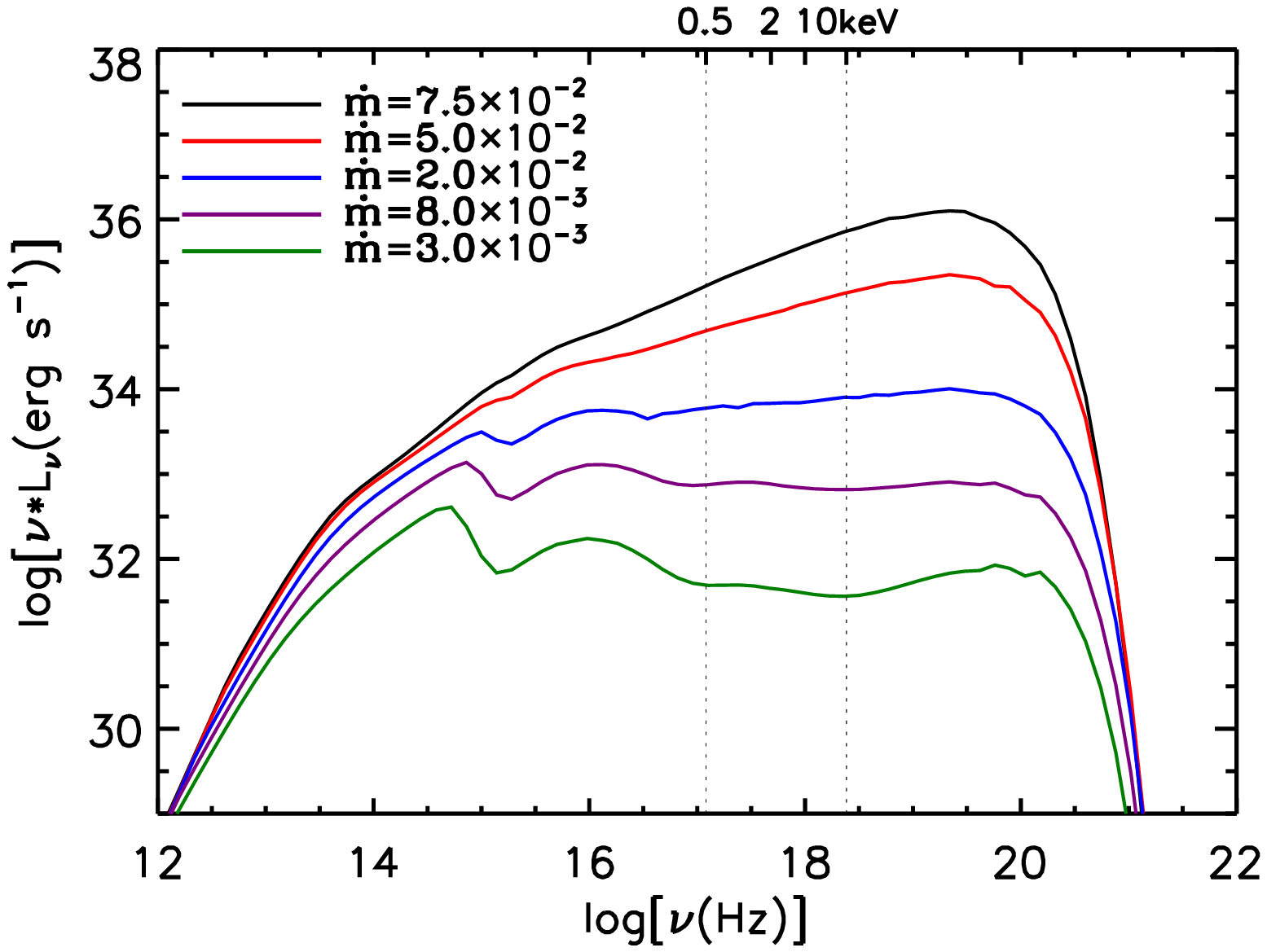}
\caption{\label{f:spbh}Emergent spectra of the ADAF around a BH. In the calculations, 
we take $m=10$, $R_{\rm in}=3R_{\rm S}$, $\alpha=1$ and $\beta=0.95$.}
\end{figure}

We plot the theoretical result of $\Gamma$ versus $L_{\rm 0.5-10\rm keV}$ based on 
the self-similar solution of the ADAF around a stellar-mass BH \citep[][]{Qiao2018b}.  
One can refer to the black filled-circle for $\Gamma$ versus $L_{\rm 0.5-10\rm keV}$ in Fig. \ref{f:gama-L}. 
The corresponding emergent spectra can be seen in Fig. \ref{f:spbh}.
In the calculation, we take BH mass $m=10$, the inner boundary of the ADAF $R_{\rm in}=3R_{\rm S}$ (with $R_{\rm S}$ 
being the Schwarzschild radius, and $R_{\rm S}=2.95\times 10^{5}\ m\ \rm cm $), viscosity parameter $\alpha=1$,
and the magnetic parameter $\beta=0.95$ as that we take for NS case. As we can see that, the 
theoretical result of $\Gamma$ versus $L_{\rm 0.5-10\rm keV}$ predicted by the ADAF around
a BH can roughly match the observational data for $\Gamma$ versus $L_{\rm 0.5-10\rm keV}$. 
We notice that the theoretical results of $\Gamma$ versus $L_{\rm 0.5-10\rm keV}$ 
for BH case is very close to that of the NS case for taking $f_{\rm th}=0$.
It is easy to understand that, if taking $f_{\rm th}=0$, it means that there is no radiative 
feedback between the surface of the NS and the ADAF, which is mathematically equivalent to the BH case that
all the energy of the ADAF crossing the event horizon of the BH will be `vanished'.
The very little difference of the theoretical results between the BH case and the
NS case for taking $f_{\rm th}=0$ is resulted by the effects of the different mass
between BHs and NSs.   

\section{Discussions}
\subsection{On the physical origin of the anti-correlation between $\Gamma$ and $L_{\rm 0.5-10keV}$}\label{s:discussion}
In this paper, we explain the observed anti-correlation between the X-ray photon index $\Gamma$ and 
the X-ray luminosity $L_{\rm 0.5-10keV}$ based a sample of non-pulsating 
NS-LMXBs in the range of $L_{\rm 0.5-10keV}\sim 10^{34}-10^{36}\ \rm erg\ s^{-1}$ 
within the framework of the self-similar solution of the ADAF around a weakly magnetized NS.
We conclude that in the range of 
$L_{\rm 0.5-10keV}\sim 10^{34}-10^{35}\ \rm erg\ s^{-1}$, the softening of the X-ray spectrum is due 
to the increase of the fractional contribution of the thermal soft X-ray component, while in the range 
of $L_{\rm 0.5-10keV}\sim 10^{35}-10^{36}\ \rm erg\ s^{-1}$, our explanation for the softening of the X-ray 
spectrum is uncertain.

As we can see from Section \ref {s:observation}, in the range of 
$L_{\rm 0.5-10keV}\sim 10^{35}-10^{36}\ \rm erg\ s^{-1}$, by comparing with
the observational data of the anti-correlation between $\Gamma$ and $L_{\rm 0.5-10keV}$ 
(i.e., equation \ref{e:fit}), it is suggested that $f_{\rm th}$ is roughly in the range of 
$0.01\lesssim f_{\rm th}\lesssim 0.1$.
In this case, the softening of the X-ray spectra with $L_{\rm 0.5-10keV}$ can be explained with a 
complex pattern, which is as follows. 
Specifically, we can see from Fig. \ref{f:gama-L} that there are three crossing points between the 
theoretical curves and equation \ref{e:fit} in the range of $L_{\rm 0.5-10keV}\sim 10^{35}-10^{36}\ \rm erg\ s^{-1}$. 
From right to left, the X-ray luminosities of the three crossing points are 
$L_{\rm 0.5-10\rm keV}\sim 1.1\times 10^{36}\ \rm erg \ s^{-1}$, 
$4.0\times 10^{35}\ \rm erg \ s^{-1}$ and $1.8\times 10^{35}\ \rm erg \ s^{-1}$ respectively,
and the theoretical value of $\eta$ of the three crossing points are 
$\eta$ $\sim 20\%$ for $f_{\rm th}=0.1$, $\sim 25\%$ for $f_{\rm th}=0.03$, and 
$\sim 40\%$ for $f_{\rm th}=0.01$ respectively as can be seen from Fig. \ref{f:eta_L_theory}.
The value of $\eta$ increases for the three crossing points with decreasing $L_{\rm 0.5-10\rm keV}$, 
which means that the fractional contribution of the thermal soft X-ray component decreases with 
decreasing $L_{\rm 0.5-10\rm keV}$. 
Here we would like to address that the effect of the thermal component on the value of $\Gamma$ 
does not only depend on the relative strength of the thermal component, but also depends on the temperature 
of the thermal component. If the temperature of the thermal component $T_{*}$ is above $\sim 0.5$ keV 
(the peak emission $T_{\rm max}$ being at $\approx 2.82T_{*}$ in $L_{\rm \nu}$ versus $\nu$), 
an increase of the strength of the thermal component will make the spectrum harder, i.e., the value of 
$\Gamma$ decreased. This is because if $T_{*}$ is above $\sim 0.5$ keV, more energy of the thermal emission 
will contribute to the hard X-rays, which intrinsically makes the spectrum harder. 
However, if $T_{*}$ is below $\sim 0.5$ keV, an increase of the strength of the thermal component will make 
the spectrum softer, i.e., the value of $\Gamma$ increased, since in this case more energy of the thermal 
emission will contribute to the soft X-rays, which intrinsically makes the spectrum softer. 
So although the fractional contribution of the thermal soft X-ray component decreases with decreasing 
$L_{\rm 0.5-10\rm keV}$, it is possible that the X-ray spectrum softens if the temperature of the
thermal component $T_{*}$ decreases from a value above $\sim 0.5$ keV to a value below $\sim 0.5$ keV. 
It is true for the three crossing points that $T_{*}$ decreases with decreasing $L_{\rm 0.5-10\rm keV}$ 
as can be easily interpolated from Table \ref{t:fth_effect} for $f_{\rm th}=0.1$, $0.03$ and $0.01$ respectively.
Meanwhile, for the three crossing points, since the value of $\eta$ is relatively small,
the evolution of $\Gamma$ is dominated by the evolution of the thermal soft X-ray component. 
So based on our model, it is suggested that the evolution of $\Gamma$ is governed by a complex relation 
between the thermal soft X-ray component and the power-law component, and the increase of $\Gamma$ 
(softening of the X-ray spectrum) with decreasing $L_{\rm 0.5-10\rm keV}$ is dominantly due to the decrease 
of the temperature of the thermal component from a value above $\sim 0.5$ keV to a value below 
$\sim 0.5$ keV.  

However, as we can see from panel (2) of Fig. \ref {f:T-frac}, in the range of 
$L_{\rm 0.5-10keV}\sim 10^{35}-10^{36}\ \rm erg\ s^{-1}$, the X-ray spectra of the sources in our sample
are dominant by a power-law component. Some other observations may also support such a scenario.   
For example, \citet[][]{Weng2015} studied the X-ray spectral evolution of the NS X-ray transient XTE J1810-189 
in 2008 outburst decay, showing that the X-ray spectrum of the 
$\it RXTE/PCA$ data (between 3-25 keV) is dominant by a power-law component ($\eta \gtrsim 90\%$) in the 
range of the X-ray luminosity of $4\times 10^{36}\ \rm erg\ s^{-1}$ to $6\times 10^{35}\ \rm erg\ s^{-1}$,
and the X-ray spectra soften with decreasing the X-ray luminosity from  
$4\times 10^{36}\ \rm erg\ s^{-1}$ to $6\times 10^{35}\ \rm erg\ s^{-1}$.
Further, in \citet[][]{Weng2015}, the authors confirmed that  
the softening of the X-ray spectrum (between 3-25 keV) with decreasing the X-ray luminosity from  
$4\times 10^{36}\ \rm erg\ s^{-1}$ to $6\times 10^{35}\ \rm erg\ s^{-1}$ is due to the softening of the 
intrinsic power-law component itself rather than the increase of the thermal soft X-ray component.
However, here we would like to mention that since $\it RXTE/PCA$ data do not cover the spectrum below 3 keV, 
it is not very clear how the spectra evolve if the soft X-ray data between 0.5 and 3 keV 
are considered. 

Finally, we would like to mention that if it is universal that in the range of 
$L_{\rm 0.5-10keV}\sim 10^{35}-10^{36}\ \rm erg\ s^{-1}$ the X-ray spectrum between
0.5 and 10 keV is dominant by a power-law component, we suggest that 
a small value of $f_{\rm th}$, i.e., $0.003\lesssim f_{\rm th}\lesssim 0.005$, still holds
in the range of $L_{\rm 0.5-10keV}\sim 10^{35}-10^{36}\ \rm erg\ s^{-1}$ 
as in the range of $L_{\rm 0.5-10keV}\sim 10^{34}-10^{35}\ \rm erg\ s^{-1}$ for explaining the observation.
For example, if we take $f_{\rm th}=0.003$, the X-ray spectrum between 
0.5 and 10 keV is nearly completely dominant by a power-law component in the range of 
$L_{\rm 0.5-10keV}\sim 10^{35}-10^{36}\ \rm erg\ s^{-1}$. In this case, the softening of the X-ray 
spectrum is dominantly due to the softening of the power-law component itself.
One can refer to the panel (5) 
in Fig. \ref {f:fth_sp} for clarity or Table \ref{t:fth_effect} for
the detailed numerical results. However, as can be seen in Fig. \ref {f:gama-L}, the theoretical relation 
of $\Gamma$ versus $L_{\rm 0.5-10keV}$ for $f_{\rm th}=0.003$ is a little deviated 
from the observations, i.e., the theoretical value of $\Gamma$ is systemically below equation \ref{e:fit} 
in the range of $L_{\rm 0.5-10keV}\sim 10^{35}-10^{36}\ \rm erg\ s^{-1}$.
We suggest that the difference between the observational data and the model predictions for 
$f_{\rm th}=0.003$ may be related with both the observations (i.e., more precise 
observational data are necessary to confirmed the relation between $\Gamma$ and $L_{\rm 0.5-10keV}$ in 
this luminosity range in the future) and the model.
Especially, in this paper, we calculate the structure and corresponding emergent spectrum of the ADAF within
the framework of the self-similar solution, and do not consider the outflow of the ADAF.
We can expect that at least if the outflow (wind) of the ADAF is considered, the 
spectrum will systematically become softer, consequently better matching the observations 
in this luminosity range. 
Here we would like to mention that recent numerical simulations of the hot accretion flow 
(i.e., ADAF) have shown that outflow is indeed existed around NSs \citep[][]{Bu2020}, as that of around 
BHs \citep[][]{Yuan2012a,Yuan2012b}. In general, numerical simulations of the ADAF show that the inflow
mass rate $\dot m(r) \propto r^{s}$, where the index $s$ is roughly in the range of 0.5-1.
That is to say if the outflow is considered, the distribution of $\dot m$ in radial direction will be
changed, especially $\dot m$ in the inner region will decrease significantly. The emission from the inner 
region of the ADAF mainly contributes to hard X-rays. A decrease of $\dot m$ in the inner region will make 
the emission in the hard X-ray band decreased, consequently making the X-ray spectrum become softer.  

\subsection{On the NSs with very hard X-ray spectra in the range of 
$L_{\rm 0.5-10keV}\sim 10^{34}-10^{36}\ \rm erg\ s^{-1}$} \label{s:hard}
Observationally, an anti-correlation between $\Gamma$ and $L_{\rm 0.5-10keV}$  
in the range of $L_{\rm 0.5-10keV}\sim 10^{34}-10^{36}\ \rm erg\ s^{-1}$
in NS-LMXBs has been proposed in \citep[][]{Wijnands2015}, which is further confirmed 
in \citet[][]{Parikh2017} and in the present paper by adding some more sources in the sample.
Meanwhile it is suggested that it is very possible that such an anti-correlation between 
$\Gamma$ and $L_{\rm 0.5-10keV}$ is `universal' for most NS-LMXBs 
in the range of $L_{\rm 0.5-10keV}\sim 10^{34}-10^{36}\ \rm erg\ s^{-1}$
\citep[][]{Wijnands2015,Parikh2017}.
However, we should note that some very hard state source in the range of 
$L_{\rm 0.5-10keV}\sim 10^{34}-10^{36}\ \rm erg\ s^{-1}$, such as IGR J17361-4441 indeed
dose not observe such a so-called `universal' anti-correlation.
Actually, a class of NSs with very hard X-ray spectra have been identified 
\citep[][]{Parikh2017}. In \citet[][]{Parikh2017}, the authors studied the X-ray spectra of six 
NS X-ray binaries, i.e., 1RXS J180408.9-342058, EXO 1745-248, IGR J18245-2452, SAX J1748.9-2021, 
IGR J17361-4441 and SAX J1808.4-3658, it is found that four out of the six sources, i.e., 
1RXS J180408.9-342058, EXO 1745-248, IGR J18245-2452 and IGR J17361-4441 show very hard X-ray spectra 
with $\Gamma\sim 1$ if the best-fitting column density $N_{\rm H}$ is used in the spectral 
fitting. IGR J18245-2452 is an accreting millisecond X-ray pulsar (AMXP), so the very hard X-ray 
spectrum is likely to be related with the stronger magnetic field. 
The two sources, 1RXS J180408.9-342058 and EXO 1745-248 show very hard X-ray spectra 
in the range of $L_{\rm 0.5-10keV}\sim 10^{36}-10^{37}\ \rm erg\ s^{-1}$. 

For IGR J17361-4441, as we can see from Fig. \ref {f:gama-L} (dark blue `$\diamond$'), in the range of    
$L_{\rm 0.5-10keV}\sim 10^{35}-10^{36}\ \rm erg\ s^{-1}$, the relation of $\Gamma$ versus $L_{\rm 0.5-10keV}$ 
significantly deviates from the so-called `universal' anti-correlation. 
IGR J17361-4441 shows very hard X-ray spectra with $\Gamma \sim$ 0.5-1.5.
Such an evolutionary pattern of $\Gamma$ with $L_{\rm 0.5-10keV}$, and the physical origin of the very 
hard X-ray spectra observed in the range of $L_{\rm 0.5-10keV}\sim 10^{35}-10^{36}\ \rm erg\ s^{-1}$
for IGR J17361-4441 are unclear, and needed to studied in detail in the future.
Here we would like to mention that although the nature of IGR J17361-4441 as 
a NS-LMXB can not be discarded \citep[][for discussions]{Wijnands2015}, alternatively, 
IGR J17361-4441 is once explained as a tidal disruption event of a planet sized body by a white dwarf 
\citep[][]{DelSanto2014}.

\subsection{The anti-correlation between $\Gamma$ and $L_{\rm 0.5-10keV}$ can be extended
below $\sim 10^{34}\ \rm erg \ s^{-1}$ and above $\sim 10^{36}\ \rm erg \ s^{-1}$? }\label{s:extended}
In this paper, we focus on the anti-correlation between the X-ray photon index $\Gamma$ and the 
X-ray luminosity $L_{\rm 0.5-10keV}$ in NS-LMXBs in the range of 
$L_{\rm 0.5-10keV}\sim 10^{34}-10^{36}\ \rm erg\ s^{-1}$. Observationally, as has been discussed 
in \citet[][]{Wijnands2015} that although there is scatter for some individual sources, in general, 
it has been demonstrated that there is a universal anti-correlation between $\Gamma$ and 
$L_{\rm 0.5-10keV}$ in the range of $L_{\rm 0.5-10keV}\sim 10^{34}-10^{36}\ \rm erg\ s^{-1}$.
However, whether the anti-correlation between $\Gamma$ and $L_{\rm 0.5-10keV}$ 
can be extended below $\sim 10^{34}\ \rm erg \ s^{-1}$ 
and above $\sim 10^{36}\ \rm erg \ s^{-1}$ are uncertain.

When the X-ray luminosity is below $\sim 10^{34}\ \rm erg \ s^{-1}$, NS-LMXBs are often 
regarded as in the quiescent state, during which the X-ray spectra are very complex and diverse. 
The X-ray spectra in the quiescent state can be, 
(1) completely dominated by a thermal soft X-ray component, or (2) completely dominated by a
power-law component, or (3) described by the two-component model, i.e., a thermal soft X-ray component
plus a power-law component \citep[][for review]{Wijnands2017}.
We notice a recent paper by \citet[][]{Sonbas2018}, in which the authors compiled a sample composed of 
twelve non-pulsating NS-LMXBs (twelve data points) in the range of 
$L_{\rm 0.5-10keV}\sim 10^{32}-10^{36}\ \rm erg\ s^{-1}$. Six data points in the sample fall in the 
range of $L_{\rm 0.5-10keV}\sim 10^{32}-10^{33}\ \rm erg\ s^{-1}$, and the X-ray spectra
can be satisfactorily described by a single power law. Finally, it is found that
there is an anti-correlation between the X-ray photon index $\Gamma$ and the X-ray luminosity  
$L_{\rm 0.5-10keV}$, and the slope of the anti-correlation is $-2.12\pm 0.63$, which 
is obviously steeper than the slope of $-0.43$ (equation \ref{e:fit}) in the range of 
$L_{\rm 0.5-10keV}\sim 10^{34}-10^{36}\ \rm erg\ s^{-1}$. 
In the range of $L_{\rm 0.5-10keV}\sim 10^{32}-10^{33}\ \rm erg\ s^{-1}$, the X-ray 
spectra of the sources in the sample of \citet[][]{Sonbas2018} are very soft. 
$\Gamma$ increases from $\sim 3.87$ to $\sim 5.8$ with $L_{\rm 0.5-10keV}$
decreasing from $\sim 10^{33}\ \rm erg \ s^{-1}$ to $\sim 10^{32}\ \rm erg \ s^{-1}$. 
Although in \citet[][]{Sonbas2018}, the authors proposed a steeper anti-correlation between 
$\Gamma$ and $L_{\rm 0.5-10keV}$ for $L_{\rm 0.5-10keV}\lesssim 10^{34}\ \rm erg \ s^{-1}$,
we think that the relation of $\Gamma$ versus $L_{\rm 0.5-10keV}$ for 
$L_{\rm 0.5-10keV}\lesssim 10^{34}\ \rm erg \ s^{-1}$ is still uncertain, which strongly depends on 
the sample selections (i.e., which kinds of X-ray spectra selected).
The physical origin of the X-ray spectra for $L_{\rm 0.5-10keV}\lesssim 10^{34}\ \rm erg \ s^{-1}$
is also uncertain, which could be dominated by crust cooling \citep[e.g.][]{Brown1998}, 
magnetospheric emission \citep[e.g.][]{Stella1994,Campana2002,Burderi2003}, or
low-level accretion (e.g., ADAF) onto NSs \citep[e.g.][]{Zampieri1995,Qiao2020}.
In our opinion, it is very necessary to fit the high quality X-ray spectra of NS-LMXBs for 
$L_{\rm 0.5-10keV}\lesssim 10^{34}\ \rm erg \ s^{-1}$ in detail to distinguish the different physical
mechanisms, which however exceeds the scope of the present paper. 
In a word, we think that the relation of $\Gamma$ versus $L_{\rm 0.5-10keV}$ for 
$L_{\rm 0.5-10keV}$ below $\sim 10^{34}\ \rm erg \ s^{-1}$ is still needed to be studied 
in detail in the future. 

When the X-ray luminosity is above $\sim 10^{36}\ \rm erg \ s^{-1}$, e.g. in the 
range of $L_{\rm 0.5-10keV}\sim 10^{36}-10^{37}\ \rm erg\ s^{-1}$, in general,
the X-ray photon index $\Gamma$ is $\sim 1.5-2$ as NS-LMXBs in the typical hard state
\citep[e.g.][]{Degenaar2012,Bahramian2014}. 
However, \citet[][]{Parikh2017} reported on unusually very hard spectral state in three NS-LMXBs 
1RXS J180408.9-342058, EXO 1745-248 and IGR J18245-2452 in the range 
$L_{\rm 0.5-10keV}\sim 10^{36}-10^{37}\ \rm erg\ s^{-1}$. 
Specifically, by fitting the {\it Swift} X-ray spectra of these three sources between 0.5 and 10 keV 
with a single power law, the authors found that the X-ray photon index of these three sources is  
very low, i.e., $\Gamma \sim0.5-1$, in the range of $L_{\rm 0.5-10keV}\sim 10^{36}-10^{37}\ \rm erg\ s^{-1}$.
The identification of the unusually hard X-ray spectra makes the relation between $\Gamma$ and 
$L_{\rm 0.5-10keV}$ in the range of $L_{\rm 0.5-10keV}\sim 10^{36}-10^{37}\ \rm erg\ s^{-1}$ very complicated. 
So if the anti-correlation correlation proposed in \citet[][]{Wijnands2015} (also in this paper) can be 
extended to the range of $L_{\rm 0.5-10keV}\sim 10^{36}-10^{37}\ \rm erg\ s^{-1}$, the scatter is very large. 
As discussed in \citet[][]{Parikh2017}, the identified unusually hard X-ray 
spectra may represent a new distinct spectral state. If such a very hard X-ray spectrum can be explained with 
the typical Comptonization model, it requires a higher electron temperature or a higher Compton scattering 
optical depth compared with the typical hard state of NS-LMXBs in this luminosity range. The physical origin for 
the higher electron temperature or the higher Compton scattering optical depth for producing the X-ray 
photon index of $\Gamma \sim 0.5-1$ in this luminosity range is still needed to be studied in the 
future in detail. 

\subsection{On the value of $f_{\rm th}$ and the radiative efficiency of 
weakly magnetized NSs with an ADAF accretion}
In our model for the ADAF accretion around a weakly magnetized NS, there is a key parameter, 
$f_{\rm th}$, describing the fraction of the ADAF energy released at the surface of the NS 
as thermal emission to be scattered in the ADAF. The value of 
$f_{\rm th}$ is very important for determining the feedback between the NS and the ADAF, 
consequently affecting the radiative efficiency of the ADAF accretion around a NS. 
However, physically, due to our relatively poor knowledge on the interaction between the NS and the 
accretion flow under the extreme gravitational field of NS, the value of $f_{\rm th}$ is uncertain,
which is probable to be related with the state of matter, the
magnetic field, as well as the thermodynamics of the accretion flow at the surface of the NS
\citep[][for discussions, and the references therein]{Wijnands2015}.  

As we have mentioned previously, in our model of the ADAF accretion, basically, there are two 
components for the X-ray spectrum between 0.5 and 10 keV, i.e., a thermal soft X-ray component and a 
power-law component. In a companion paper \citet[][]{Qiao2020}, as the 
zeroth order approximation for testing the model predictions, we theoretically investigate the 
correlation between the fractional contribution of the power-law component $\eta$ and the X-ray luminosity 
$L_{\rm 0.5-10keV}$ for a sample of NS-LMXBs probably dominated by low-level accretion onto NSs
in a wider X-ray luminosity range from 
$L_{\rm 0.5-10keV}\sim 10^{32}\ \rm erg\ s^{-1}$ to $\sim 10^{36}\ \rm erg\ s^{-1}$. 
It is found that a small value of $f_{\rm th}$, i.e., $f_{\rm th}\lesssim 0.1$ is needed to match the 
observed correlation between $\eta$ and $L_{\rm 0.5-10keV}$. 
In this paper, we further test the model predictions by explaining the observed 
$\Gamma-L_{\rm 0.5-10keV}$ anti-correlation based on a sample of non-pulsating 
NS-LMXBs in the range of $L_{\rm 0.5-10keV}\sim 10^{34}-10^{36}\ \rm erg\ s^{-1}$.
We conclude that $f_{\rm th}$ is between 0.003 and 0.1, which further confirms the previous conclusion
in \citet[][]{Qiao2020}, i.e., $f_{\rm th}\lesssim 0.1$. 
The small value of $f_{\rm th}\lesssim 0.1$ from \citet[][]{Qiao2020} 
and the present paper jointly suggests that the radiative efficiency of 
weakly magnetized NSs with an ADAF accretion is not as high as the generally proposed value of
$\epsilon \sim {\dot M GM\over R_{*}}/{\dot M c^2}\sim 0.2$. 
We would like to mention that there is a very interesting paper  
\citet[][]{DAngelo2015}, in which the authors investigated the radiative efficiency
of a NS X-ray transient Cen X-4 by fitting its broad band spectrum 
at a luminosity of $L_{\rm X}\sim 10^{33}\ \rm erg\ s^{-1}$.
In general, our ADAF model predicts a similar radiative efficiency as that of 
\citet[][]{DAngelo2015}, one can refer to Section 4.3 in \citet[][]{Qiao2020} for 
detailed discussions. 
As discussed in Section 4.3 of \citet[][]{Qiao2020}, it is very possible that the remaining fraction, 
i.e., 1-$f_{\rm th}$, of the ADAF energy transferred onto the surface of the NS could be converted to the 
rotational energy of the NS.  

\section{Conclusions}
In this work, we explain the observed anti-correlation between the X-ray photon index 
$\Gamma$ (obtained by fitting the X-ray spectrum between 0.5 and 10 keV with a single power law) 
and the X-ray luminosity $L_{\rm 0.5-10keV}$ in NS-LMXBs in the range of 
$L_{\rm 0.5-10keV}\sim 10^{34}-10^{36}\ \rm erg\ s^{-1}$ within the framework of the self-similar 
solution of the ADAF around a weakly magnetized NS.
We show that the ADAF model intrinsically can predict an anti-correlation between  
$\Gamma$ and $L_{\rm 0.5-10keV}$. We test the effect of a key parameter, $f_{\rm th}$, describing the 
fraction of the ADAF energy released at the surface of the NS as thermal emission to be scattered in 
the ADAF, on the anti-correlation between $\Gamma$ and $L_{\rm 0.5-10keV}$. We found that the value of 
$f_{\rm th}$ can significantly affect the slope of the $\Gamma-L_{\rm 0.5-10keV}$ anti-correlation.
Specifically, the anti-correlation between $\Gamma$ and $L_{\rm 0.5-10keV}$
becomes flatter with decreasing $f_{\rm th}$ as taking $f_{\rm th}=0.1, 0.03, 0.01, 0.005$, $0.003$ and $0$ 
respectively. 
By comparing with a sample of non-pulsating NS-LMXBs with well measured $\Gamma$ and $L_{\rm 0.5-10keV}$, 
it is found that the value of $0.003\lesssim f_{\rm th}\lesssim 0.1$ is needed to match the observed 
$\Gamma-L_{\rm 0.5-10keV}$ anti-correlation. Finally, we argue that the small value of 
$f_{\rm th}\lesssim 0.1$ derived in this paper further confirms the previous conclusion that the 
radiative efficiency of NSs with an ADAF accretion may not be as high as 
$\epsilon \sim {\dot M GM\over R_{*}}/{\dot M c^2}\sim 0.2$ as proposed in \citet[][]{Qiao2020}. 
As for the remaining fraction, i.e., 1-$f_{\rm th}$, of the ADAF energy transferred onto the surface of 
the NS, we suggest that one of the promising possibilities is that such energy could be converted to the 
rotational energy of the NS, the test of which is still needed in the future.

\section*{Acknowledgments}
This work is supported by the National Natural Science Foundation of 
China (Grants 11773037 and 11673026), the gravitational wave pilot B (Grants No. XDB23040100), 
the Strategic Pioneer Program on Space Science, Chinese Academy of Sciences 
(Grant No. XDA15052100) and the National Program on Key Research and Development 
Project (Grant No. 2016YFA0400804).

\bibliographystyle{mnras}
\bibliography{qiaoel}


\bsp	
\label{lastpage}
\end{document}